\documentclass[twocolumn]{aastex701}

\usepackage{comment}
\usepackage{xspace}

\newcommand{\exocomp}{\texttt{ExoComp}\xspace}

\newcommand{\jwst}{\textit{JWST}\xspace}

\newcommand{\hst}{\textit{HST}\xspace}

\newcommand{\spit}{\textit{Spitzer}\xspace}

\newenvironment{my_itemize}{
	\begin{itemize}
		\setlength{\itemsep}{1pt}
		\setlength{\parskip}{0pt}
		\setlength{\parsep}{0pt}}{\end{itemize}
}
 
\begin{document}

\title{The Library of Exoplanet Atmospheric Composition Measurements: Population Level Trends in Exoplanet Composition with \exocomp{}}

\author[orcid=0000-0003-3667-8633]{Joshua D. Lothringer}
\affiliation{Space Telescope Science Institute, Baltimore, MD, USA}
\email[show]{jlothringer@stsci.edu}  

\author[orcid=0000-0001-6508-5736]{Nataliea Lowson}
\affiliation{Department of Physics and Astronomy, University of Delaware, 217 Sharp Lab, Newark, DE 19716, USA}
\email[]{nlowson@udel.edu} 

\author[orcid=0000-0002-3263-2251]{Guangwei Fu}
\affiliation{Department of Physics and Astronomy, Johns Hopkins University, Baltimore, MD, USA}
\email[]{nlowson@udel.edu}  

\begin{abstract}

The present-day bulk elemental composition of an exoplanet can provide insight into a planet's formation and evolutionary history. Such information is now being measured for dozens of planets with state-of-the-art facilities using Bayesian atmosphere retrievals. We collect measurements of exoplanet composition of gas giants into a Library of Exoplanet Atmospheric Composition Measurements for comparison on a population level. We develop an open-source toolkit, \exocomp{}, to standardize between solar abundance, metallicity, and C/O ratio definitions. We find a systematic enhancement in the metallicity of exoplanets compared to T-dwarf and stellar populations, a strict bound in C/O between 0 and 1, and statistically significant differences between measurements from direct, eclipse, and transmission spectroscopy. In particular, the transit spectroscopy population exhibits a systematically lower C/O ratio compared to planets observed with eclipse and direct spectroscopy. While such differences may be astrophysical signals, we discuss many of the challenges and subtleties of such a comparison. We characterize the mass-metallicity trend, finding a slope consistent between planets measured in transit versus eclipse, but offset in metallicity. Compared to the Solar System and constraints from interior modeling, gas giant atmospheres appear to exhibit a steeper mass-metallicity trend. We hope that the tools available in \exocomp{} and the data in the Library of Exoplanet Atmospheric Composition Measurements can enhance the science return of the wide-array of space- and ground-based exoplanet science being undertaken by the community.

\end{abstract}

\received{2025 September 29}

\revised{2025 October 27}

\accepted{2025 October 30}


\setcounter{footnote}{0}
\section{Introduction}

One of the primary ways in which we study planets is through their composition. Measurements of atomic and molecular abundances can be used to study thermochemical processes like mixing, photochemistry, and molecular dissociation. However, a fundamental motivation is to connect a planet's present day composition to the building blocks from which a planet formed, as this can reveal information about the planet's formation, migration, and evolutionary history. 

The bulk metallicity of an exoplanet can be enhanced above the host star's primordial metallicity through the accretion of rocky and/or icy material \citep{pollack:1996}, or through the accretion of gas enriched in heavy elements, such as at evaporation fronts inside of ice lines \citep[e.g.,][]{oberg:2016}. The abundance of carbon relative to oxygen, i.e., the C/O ratio, can provide insight into the composition of the building blocks that a planet accreted \citep[][]{oberg:2011}. For example, the accretion of H$_2$O ice could lower a planet's C/O ratio, while the accretion of gas at the same location would increase the C/O ratio due to the sequestration of oxygen in the solid phase. 

It has now been recognized that the C/O ratio alone is insufficient information to trace the complex, dynamic, and stochastic process of planet formation \citep[e.g.,][]{mordasini:2016,molliere:2022,feinstein:2025}. Many different migration and accretion scenarios (e.g., gas- versus solid-enrichment) make the C/O ratio degenerate with respect to formation history. More refractory species can help break these degeneracies by specifically tracing the rocky building blocks of a planet. Sodium and potassium \citep[][]{welbanks:2019a}, sulfur \citep[][]{crossfield:2023}, and heavy metals like Fe and Mg \citep[][]{lothringer:2021} have all been identified as promising tracers in this regard. By then measuring a refractory-to-volatile (Refr./Vol.) abundance ratio, we can infer the relative proportion of rocky and icy material that enriched a planets atmosphere \citep{lothringer:2021,turrini:2021}.

In the new era enabled by \jwst{} \citep{rigby:2022} and new instruments at 8 meter-class ground-based facilities, we have access to more reliable measurements of a wider array of atmospheric species than ever before~\citep[][]{espinoza:2025,feinstein:2025}. Measurements of H$_2$O, CO, CO$_2$, and/or CH$_4$ are now regularly available to constrain the volatile enrichment and C/O ratio. A multitude of refractory species, from Na and K to Fe and Mg, have also been measured from the UV to IR \citep[][and references therein]{feinstein:2025}, providing insight into the rocky component of planet enrichment.

Bayesian atmospheric retrievals have become a widely-used method for measuring a planet's composition~\cite[e.g.,][]{madhusudhan:2009,line:2012,benneke:2012,madhu:2018review}, yet disparate techniques have hindered the inter-comparison of results~\citep[for an overview, see][]{barstow:2020b}. While previous works have focused on technical aspects like molecular line list selection, other differences in retrieval codes can hinder inter-comparison. For example, the seemingly straightforward choice of a reference solar abundance can lead to differences in the metallicity and C/O ratio. In the worst case scenario, differences in the definition of metallicity via [O/H] can be over 17\% different between \cite{asplund:2021} and \cite{lodders:2025}. Even worse, the C/O ratio is over 22\% different between \cite{lodders:2010} and \cite{asplund:2021}. Additionally, different studies' definitions of metallicity \citep{heng:2018} and C/O hinders inter-comparison.

As a step towards a homogeneous comparison of disparate retrieval results, we introduce \exocomp{}, a toolkit with utilities for comparing measurements of the atmospheric composition of exoplanets. The tools in \exocomp{} include:
\begin{enumerate}
    \item \texttt{convert\_solar}, for changing the underlying solar abundance definitions to a different source (i.e., \cite{asplund:2021} to \cite{lodders:2025} or vice-versa.
    \item \texttt{define\_stellar}, enables a user to define their own set of reference primordial abundances based on host star measurements \citep[e.g.,][]{polanski:2022,reggiani:2022,reggiani:2024}.
    \item \texttt{MMR\_to\_VMR} and \texttt{VMR\_to\_MMR}, for conversions between mass fractions and volume mixing ratios, two common but different ways of representing the abundance in atmospheric retrievals.
    \item \texttt{convert\_bulk\_abundance}, which converts metallicity, C/O, and/or Refr./Vol. to absolute elemental abundances, taking into account the appropriate solar abundance, metallicity, and C/O definitions.
    \item \texttt{convert\_species\_abunds}, which fits metallicity, C/O, and/or Refr./Vol. to a set of observed species abundances (e.g., H$_2$O, CO$_2$, CH$_4$) as an additional way to constrain bulk abundances from free retrievals.
\end{enumerate}

Here, we use \exocomp{} to standardize and compare metallicities and C/O ratios compiled from chemical equilibrium retrievals of data from \jwst{} and/or 8-m-class telescopes. Providing a homogeneous data set of compositional measurements through \exocomp{} enables a basis for population-level comparison. Such a comparison (with 65 individual measurements of 46 unique planets as of this publication) has the statistical power to robustly identify trends in bulk composition versus planet properties. Additionally, we can also use this dataset to identify potential biases in our measurement techniques. For example, differences in bulk abundance inferences between planets measured in different geometries (e.g., transit versus eclipse) may point to the influence of biases like limb-limb asymmetries or non-uniform dayside hemispheres.

In what follows, we define ``bulk abundance" to mean quantities like the metallicity, C/O ratio, or refractory-to-volatile (Refr./Vol.) ratio. We take metallicity to mean the overall enrichment of all elements heavier than hydrogen and helium, quoted in number fraction. ``Elemental abundances" then refers to the abundance of specific elements relative to hydrogen (e.g., O/H). In some cases, we quote elemental abundances in bracket notation (e.g., [O/H]), which conventionally refers to the base-10 logarithm of the oxygen-to-hydrogen ratio relative to the corresponding solar abundances:
\begin{equation}
    \mathrm{[O/H]} = \log_{10} \frac{\mathrm{O/H}}{\mathrm{O_{\odot}/H_{\odot}}}.
\end{equation}
\noindent Lastly, ``species abundance" refers to the abundance of individual atoms or molecules (e.g., H$_2$O, CO, etc.). By default, we will refer to all abundances in number-ratio or volume mixing ratio, but we will also discuss abundances in terms of mass fractions as well.

The rest of this work is organized into three sections. In Section~\ref{sec:exocomp} we introduce and provide an overview to the \exocomp{} package, which is then applied to collect a standardized sample of 66 Jovian exoplanet measurements in Section~\ref{sec:table_analysis}. We first look at trends in planetary metallicity in Section~\ref{sec:met}, including the mass-metallicity trend (Section~\ref{sec:mass_met}). We then discuss trends in the C/O ratio in Section~\ref{sec:co}. We explore the relationship between metallicity and C/O with stellar properties in Section~\ref{sec:stellar} before concluding our discussion in Section~\ref{sec:conclude}. All calculations in this paper are replicated in Jupyter Notebooks as part of the documentation for \exocomp{} (see Footnotes 4 and 5) with the version of the software and notebooks used here archived in Zenodo \citep{exocomp_zenodo}.

\section{\texttt{ExoComp}}
\label{sec:exocomp}
We introduce The Exoplanet Composition Toolkit, or \exocomp{}, a publicly available, open-source set of utilities for the inter-comparison of exoplanet composition measurements. Written in Python and available on Github\footnote{https://github.com/jlothringer/exocomp}, \exocomp{} uses a class-based framework where sets of abundances are associated with attributes necessary for inference and inter-comparison, namely the solar abundance reference, the specific metallicity definition ([O/H], [C/H], [(O+C)/H], etc.), and the parameterization of C/O (changing the O or C abundance, or both). This allows users to convert easily between definitions and to convert to elemental ratios (e.g., O/H). \exocomp{} is integrated with proper error propagation (including the errors in the solar abundance definitions), which can be especially tricky when including upper- or lower-limits and/or asymmetric uncertainties. The package also enables the flexibility to input sampled posteriors from retrievals rather than defining errors. We refer the reader to live documentation\footnote{https://exocomp.readthedocs.io/en/latest/} for instructions on installation and use, however, we detail below the major utilities within \exocomp{}, outlining their motivation, implementation, and limitations.

\subsection{Defining Reference Abundances}

There currently exists several in-use definitions of solar elemental abundances, reflecting our on-going knowledge of our Solar System's primordial composition. Unfortunately, this can create an ambiguity when measurements are simply quoted as ``solar" without reference to a specific definition. As stated above, definitions of solar metallicity and C/O in widely-used references can vary at the tens of percent level \citep[e.g.,][]{lodders:2010,asplund:2021,lodders:2025}, hindering our ability to discern astrophysical trends or measurement biases. 

Fortunately, as long as the solar abundance reference is defined, we can easily convert either to elemental abundances (e.g., O/H, C/H, etc.) or to other reference systems. \exocomp{} has tabulated solar abundance definitions from \cite{asplund:2005:abund,asplund:2009,lodders:2010,caffau:2011,lodders:2021,asplund:2021,lodders:2025}. Users can provide at input the solar abundance definition being used. Alternatively, users can label their input with the retrieval used, for which the associated solar definition is tabulated. Table~\ref{tab:retrieval_codes} lists the current definitions. With this information, users can then convert metallicities and C/O ratios in one system to another, or convert to elemental abundances. The solar abundance definition and composition parameterizations that each retrieval code uses can change over time, so we caution users to be sure the associated solar abundance definitions are correct for their dataset.

\subsubsection{Defining Stellar Abundances: WASP-77Ab Case-Study}

When attempting to infer the formation history of an individual planet, it will often not suffice to simply compare to solar abundances, since there is no guarantee that solar elemental abundances are an adequate representation of a given planetary system's primordial abundances. Rather, the proper comparison would be to the elemental abundances measured from the host star. While not all planet host stars have full chemical inventories, such information should be incorporated where available.

Here, we show a recent example with WASP-77Ab measured at at high-resolution with Gemini/IGRINS ($R\sim 45,000$) \citep{line:2021} and \jwst{}/NIRSpec ($R\sim 2,700$\footnote{\jwst{} defines $R\sim 2,700$ as high-resolution, $R\sim1000$ as medium resolution and $R\sim 100$ as low resolution. However, we acknowledge that when incorporating ground-based spectrographs, $R\gtrsim25,000$ is usually considered as high-resolution and $1000\lesssim R < 25,000$ is considered as medium-resolution.}) \citep{august:2023}. We take the values relative to solar from the literature and update them using \exocomp{} to obtain the metallicity and C/O ratio relative to the measured stellar abundances from \cite{reggiani:2022}. 

The high-resolution observations measured a solar-like C/O ratio of $0.59 \pm 0.08$ in the planet's atmosphere. \cite{reggiani:2022} subsequently measured the carbon and oxygen abundances of the host star and found a sub-solar stellar C/O ratio of $0.44^{+0.07}_{-0.08}$. These observations indicate that while the planet is only 7\% more enriched in carbon than the Sun, it is actually approximately 35\% enriched when compared to its host star. 

We note that the NIRSpec/G395H observations presented in \cite{august:2023} measure a C/O ratio of $0.36^{+0.10}_{-0.09}$ in the planet -- closer to the stellar value than the high-resolution observations. Interestingly, both the Gemini/IGRINS and \jwst{}/NIRSpec observations suggest that the planet is significantly sub-solar in overall heavy element enrichment, measuring metallicities of $-0.48$ and $-0.91$, respectively, relative to solar \citep{asplund:2009}. Here again, measurements of the host star are useful, as \cite{reggiani:2022} measures a [(C+O)/H]$=0.33 \pm 0.09$, suggesting an even more sub-stellar enrichment than one would assume from the solar elemental abundances of \cite{asplund:2009}. With \exocomp{}, we measure [C+O/H] = $-0.68 \pm 0.2$ and $-1.16 \pm 0.35$ from the Gemini/IGRINS and \jwst{}/NIRSpec observations, respectively.

\subsection{Converting from Bulk Abundances to Elemental Abundances}\label{sec:bulk_to_elemental}

Another issue exists where differences in the definition of metallicity and C/O ratio can lead to different elemental abundances when given the same bulk abundances. This is a consequence of the various methods by which the bulk abundances are parameterized in chemical equilibrium retrieval algorithms. For example, in \texttt{PETRA}, when directly measuring the C/O ratio and metallicity, all elemental abundances are first scaled to the chosen metallicity value and then the carbon abundances is varied to reach the chosen C/O ratio \citep{lothringer:2020a}. In the same scenario with \texttt{pRT} \citep{molliere:2019}, the C/O ratio is actually varied by modifying the oxygen abundance.\footnote{\url{https://petitradtrans.readthedocs.io/en/latest/content/notebooks/interpolating_chemical_equilibrium_abundances.html}} A major drawback of these methods is that the C/O is modifying the \textit{actual} heavy element fraction \textit{after} the elemental abundances have already been scaled by the metallicity and thus there is a disconnect between the metallicity parameter and the ``true" metal enrichment of the atmosphere.

A third common way to parameterize the metallicity and C/O ratio is to vary the carbon and oxygen abundance simultaneously to achieve the correct C/O ratio, while preserving the total metallicity. This can be done as follows:

\begin{equation}
    \mathrm{C/H = (C/O * ((C+O)/H)) / (C/O + 1) } \label{eq:CH}
\end{equation}
\begin{equation}
    \mathrm{O/H = ((C+O)/H) / (C/O + 1).} \label{eq:OH}
\end{equation}

\noindent With these three definitions commonly used in retrieval packages, a pair of metallicity and C/O ratios may actually correspond to three different sets of [O/H] and [C/H].

Many retrieval packages, including \texttt{PETRA} and \texttt{pRT}, have gotten around this problem (or avoided it entirely), by including the ability to retrieve the elemental abundances ratios (relative to hydrogen) directly. This is generally the recommended way to parameterize the abundances in chemical equilibrium because it enables more flexibility, especially when including additional elemental ratios (e.g., N/O) into the retrieval. The drawback is that the chemistry then needs to be calculated on-the-fly, i.e., the chemical equilibrium solution needs to be calculated every iteration since abundances are not generally pre-tabulated. By retrieving just metallicity and C/O, tables can be pre-tabulated before running the retrieval, saving computation time. 

Within \exocomp{}, measurements of metallicity and C/O are associated with the definition that was used in their retrieval. Then, in the conversion from bulk abundances to elemental abundances, the appropriate conversion is used. Table~\ref{tab:retrieval_codes} shows the default metallicity and C/O definitions for various retrieval codes. A metallicity type of (O+C)/H and a C/O type of ``\texttt{MH\_Preserve}" refers to the procedure described in Equations~\ref{eq:CH} and \ref{eq:OH}. Note again, that these definitions can change over time or may be flexible within retrieval codes, so these definitions should be clearly stated for each analysis in the literature. To match solar abundance conventions, we represent the elemental abundances as $\log_{10}$ parts per trillion (ppt), with H defined as $10^{12}$. 

\begin{deluxetable}{lcccc}
\tabletypesize{\small}
\tablecaption{Summary of metallicity and C/O treatment in various retrieval codes\label{tab:retrieval_codes}}
\tablehead{
\colhead{Code} & \colhead{[M/H] Type} & \colhead{C/O Type} & \colhead{Solar Abundances} & \colhead{Reference}
}
\startdata
\texttt{pRT} & C/H & O/H & \cite{asplund:2009} & \cite{molliere:2019}, see footnote 7\\
\texttt{PICASO} & (O+C)/H & \texttt{MH\_Preserve} & \cite{lodders:2010} & \cite{mukherjee:2023} \\
\texttt{PETRA} & O/H & C/H & \cite{asplund:2005:abund} & \cite{lothringer:2020a} \\
\texttt{POSEIDON\tablenotemark{a}} & O/H & C/H & \cite{asplund:2021} & \cite{macdonald:2023,meech:2025} \\
\texttt{CHIMERA} & (O+C)/H & \texttt{MH\_Preserve} & \cite{asplund:2009} & \cite{line:2013} \\
\texttt{ScCHIMERA} & (O+C)/H & \texttt{MH\_Preserve} & \cite{asplund:2009} & \cite{line:2013}\tablenotemark{b} \\
\texttt{SCARLET} & O/H & C/H & \cite{asplund:2009} & \cite{benneke:2015,pelletier:2025} \\
\texttt{ATMO} & (O+C)/H & \texttt{MH\_Preserve}& \cite{asplund:2009} & \cite{tremblin:2017} \\
\texttt{PLATON} & O/H & C/H & \cite{asplund:2009} & \cite{zhang:2019} \\
\texttt{PLATON 6} & (O+C)/H & \texttt{MH\_Preserve}& \cite{asplund:2009} & \cite{zhang:2025} \\
\texttt{Gibson} & (O+C)/H & \texttt{MH\_Preserve} & \cite{asplund:2009} & \citep{gibson:2022} \\
\texttt{HyDRA} & (O+C)/H & \texttt{MH\_Preserve}& \cite{asplund:2021} & \cite{gandhi:2018} \\
\enddata
\tablenotetext{a}{\texttt{POSEIDON} may vary O/H to achieve desired C/O (Meech et al. 2025).}
\tablenotetext{b}{Also see \cite{mansfield:2024}.}
\tablenotetext{}{The \texttt{MH\_Preserve} parameterization is described in Eqs~\ref{eq:CH} and \ref{eq:OH}.}
\tablenotetext{}{For retrievals that do not use pre-tabulated equilibrium abundances and retrieve O/H and C/H directly, we label their [M/H] and C/O types as (O+C)/H and \texttt{MH\_Preserve}, respectively.}
\end{deluxetable}

\subsection{Inferring Metallicity and C/O from Free Retrievals}

By freely varying the abundance of individual atmospheric species, so-called free retrievals are a flexible alternative to chemical equilibrium retrievals. Free retrievals are ideal for identifying the presence of species, and measuring atomic and molecular abundances to track processes like disequilibrium chemistry, thermal dissociation, and photochemistry. A disadvantage of a free retrieval setup is that bulk and elemental abundances are not directly constrained. Rather, a common way to measure bulk or elemental abundances from a free retrieval is to sum the number of atoms for a given element for all detected species, e.g.:

\begin{equation}\label{eq:add-up}
    \mathrm{O/H = (H_2O)_{vmr} + (CO)_{vmr} + (2*CO}_{2})_{ \mathrm{vmr}},
\end{equation}

\noindent where ``vmr" denotes the volume-mixing ratio. We refer to this as the ``Species Summation Method".

However, there are two important drawbacks of this approach. First, unless one knows that all species that contain a given element have been detected, the measurement on the elemental abundance will only be a lower limit; there may be unaccounted atoms of a given element in species that are not detected. A common example of this is the sequestration of oxygen in condensed species like silicates, which can `hide' about 20\% of oxygen atoms \citep{burrows:1999,line:2021}.

The second drawback is that large uncertainties on high-abundance species have a disproportionate effect on the subsequent uncertainty on the metallicity and C/O ratio. In effect, the uncertainty on the bulk abundance is determined by the weakest link in the chain, i.e., the species with the poorest measurement. An example is CO, which has generally weaker opacity than CO$_2$, despite being more abundant. A precise measurement of H$_2$O and CO$_2$ with NIRSpec/G395H therefore cannot be translated into a precise constraint on O/H and C/H if CO is not equally well constrained.

Here, we explore another way to infer bulk and elemental abundances from free retrieval constraints by fitting the measured species abundances to chemical equilibrium expectations. Using a chemical equilibrium solver with flexible input elemental abundances, we can fit for the elemental ratios that best explain a collection of species abundances from a free retrieval. This technique can be used to assess whether free retrieval results are consistent with chemical equilibrium retrieval results. Another key advantage of this technique is that one can infer bulk abundances without having to account for every element in every species-- one does not even need to measure the most abundant species. 

This is useful in cases where, for example, CO is not measured precisely, but is likely to be the dominant carbon carrier, if not the dominant heavy-element species. This is often the case with \jwst{} observations of hot Jupiter atmospheres \citep[e.g.,][]{rustamkulov:2022}. Instead, we can turn measurements of H$_2$O and CO$_2$, which are more readily measured, into more precise constraints on a planet's metallicity and C/O ratio without the need to measure CO. The key shortcoming of this method is its assumption of chemical equilibrium, which may not always hold in the presence of mixing, photochemistry, etc. However, one may be able to infer how far out of equilibrium a species is based on the goodness-of-fit to a chemical equilibrium solution. 

A secondary disadvantage to fitting with chemical equilibrium solvers is that one must also have an idea of the pressure and temperature at which the measured species abundance is representative. This is less of a problem for planets away from large temperature-dependent chemical transitions like the CO-CH$_4$ transition \citep{visscher:2011} or the onset of thermal dissociation \citep{parmentier:2018,lothringer:2018b,kitzmann:2018}. Additionally, if an accurate temperature and pressure are known, then this is also not an issue; for example, the measurement of the H$_2$O abundance in an ultra-hot Jupiter can still be turned into a constraint on [O/H] if an accurate temperature and pressure are known because thermal dissociation is still an equilibrium process.

With that said, we explore here the efficacy of using chemical equilibrium solvers to fit freely retrieved species abundances. We use \texttt{easyChem} \citep{lei:2024} to solve for chemical equilibrium given a set of input elemental abundances. Associated with these elemental abundances can be a set of error bars. If these error bars are symmetric, we use a least-squares algorithm (\texttt{scipy.optimize.curve\_fit}) to calculate the best-fit and uncertainty of the bulk abundances from the square-root of the diagonal of the resulting covariance matrix. If the error bars are asymmetric, we use a minimizer (\texttt{scipy.optimize.fmin}) to minimize a custom cost-function that penalizes the residuals to the fit based on the appropriate side of the errorbar. For upper-limits on a species' abundance, we only penalize a set of bulk abundances if it results in a species abundance above the measured value. Alternatively, posterior samples of the measured species abundances can be input for which a best-fit chemical equilibrium solution will be calculated for each sample, resulting in the corresponding posterior distribution of bulk abundances.

Below, we show examples of converting free retrieval estimates into bulk abundance constraints for \jwst{} observations of WASP-17b and WASP-178b.

\begin{figure}
    \centering
    \includegraphics[width=1\linewidth]{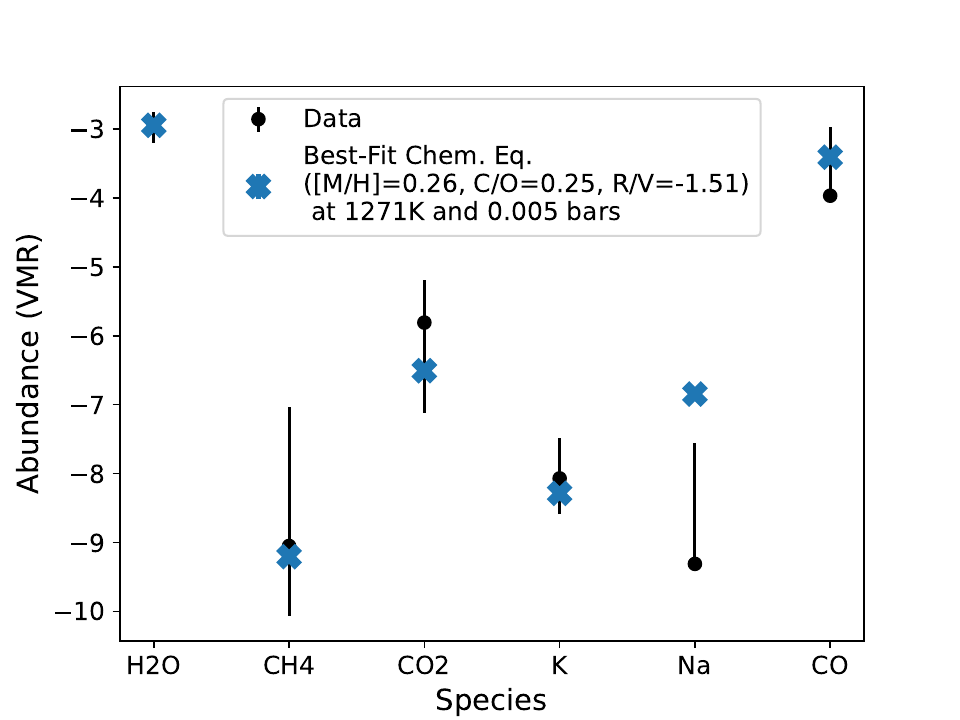}
    \caption{Measured atmospheric abundances in WASP-17b from \cite{louie:2025} (black points with error bars) compared to the best-fit chemical equilibrium solution at [M/H] = 0.26, C/O = 0.25, and R/V= $-1.51$ (blue X's). Upper-limits from Na and CO are represented by a one-sided error bar.}
    \label{fig:w17b}
\end{figure}

\begin{table}[h!]
\centering
\begin{tabular}{lcc}
\hline
\multicolumn{3}{c}{\textbf{Species Abundances}} \\
\hline
Species & Retrieved Abundance  & Best-fit Abundance \\
\hline
H$_2$O & $-2.96^{+0.21}_{-0.24}$ & $-2.95$ \\
CH$_4$ & $-9.05^{+2.01}_{-1.01}$ & $-9.22$ \\
CO$_2$ & $-5.81^{+0.62}_{-1.31}$ & $-6.49$ \\
K      & $-8.07^{+0.58}_{-0.52}$ & $-8.28$ \\
Na     & $-9.31^{+1.75}_{-0.00}$ & $-6.96$ \\
CO     & $-3.97^{+1.00}_{-0.00}$ & $-3.40$ \\
\hline
\multicolumn{3}{c}{\textbf{Inferred Bulk Composition}} \\
\hline
 & \uline{Species Summation Method} & \uline{Chem. Eq. Fit} \\
Metallicity & $0.10^{+0.41}_{-0.19}$ & $0.26\pm0.24$ \\
C/O         & $0.09^{+0.80}_{-0.09}$ & $0.25\pm0.32$ \\
\phantom{}[Refr./Vol.] & $-2.55 \pm 0.64$ & $-1.51\pm0.60$ \\
\hline
\end{tabular}
\caption{Top section: Retrieved $\log_{10}$ abundances of WASP-17b from \cite{louie:2025} using the ``Model B" \texttt{POSEIDON} free retrieval compared to our best-fit chemical equilibrium scenario. We interpret the Na and CO abundances as upper-limits. Bottom section: Resulting bulk composition inferences using the species summation method from Eq.~\ref{eq:add-up} (left) and from fitting a chemical equilibrium solution with \exocomp{} (right).\label{tab:w17b}}
\end{table}

\subsubsection{WASP-17 b}

\cite{louie:2025} recently measured abundances of H$_2$O, CH$_4$, CO$_2$, and K, with upper-limits on the abundance of Na and CO (see Table~\ref{tab:w17b}). We use these values as input into \exocomp{} as an example for converting from free retrieval results to bulk and elemental abundances. This example is particularly apt due to the poorly constrained CO, which would otherwise confound a measurement of metallicity and C/O ratio by adding up the detected species. By adding up the atoms and molecules, we estimate a metallicity of $0.10^{+0.41}_{-0.19}$ and C/O of $0.09^{+0.80}_{-0.09}$. 

If we fit a chemical equilibrium solution to the measurements, we can overcome the low-precision CO measurement by using the information from the other species to infer a metallicity and a C/O ratio. Doing so, we calculate a metallicity of $0.12\pm0.21$ and C/O of $0.15\pm0.20$. If we include the Refr./Vol. as another parameter, adjusting K and Na abundances relative to our calculated metallicity, we get a $0.26\pm0.24$, a C/O of $0.25\pm0.32$, and Refr./Vol. of $-1.51\pm0.60$.

Figure~\ref{fig:w17b} shows the measured species abundances compared to the best-fit set of abundances with the chemical equilibrium solution found above (including a refractory-to-volatile ratio). The measured abundances are well-fit by the chemical equilibrium solution, with a reduced $\chi^2$ of 0.021 (a reduced $\chi^2 << 1$ is acceptable because the species abundance measurements are not independent events with respect to measuring the corresponding bulk abundance parameters). 

\begin{figure}
    \centering
    \includegraphics[width=1\linewidth]{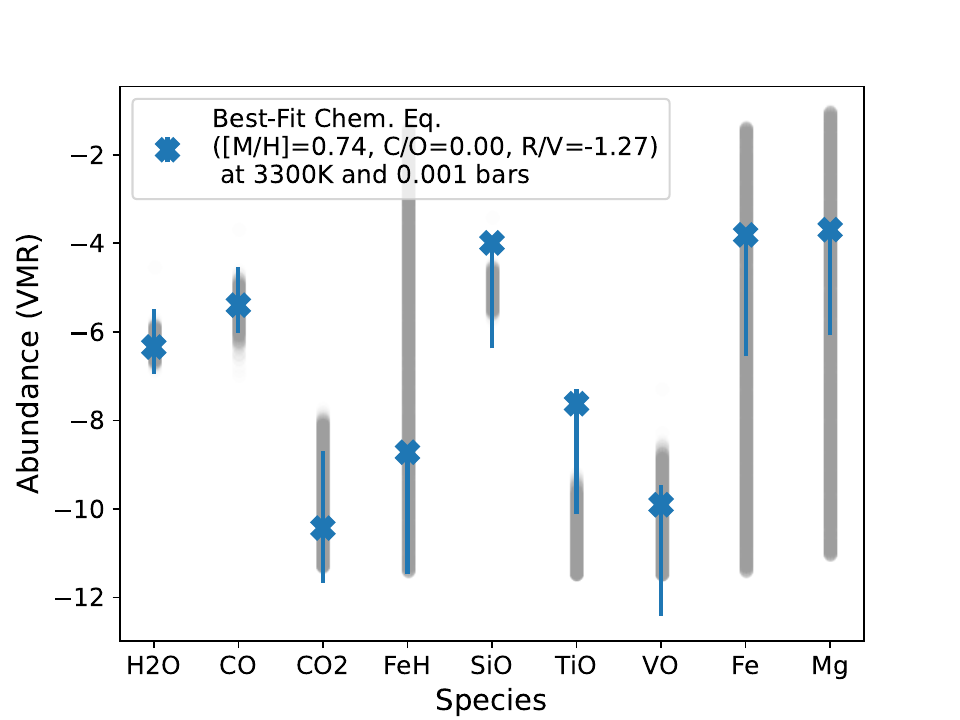}
    \caption{Measured atmospheric abundances in WASP-178b from \cite{lothringer:2025} (transparent grey points representing posterior range) compared to the best-fit chemical equilibrium solution at [M/H] = 0.70, C/O = 0.002, and R/V of $-1.33$ (blue X's). We assume a temperature of 3300~K and pressure of 0.001 bar for the chemical equilibrium fit, consistent with the retrieved isothermal temperature structure.}
    \label{fig:w178b}
\end{figure}

\begin{table}[h!]
\centering
\begin{tabular}{lcc}
\hline
\multicolumn{3}{c}{\textbf{Species Abundances}} \\
\hline
Species & Retrieved Abundance  & Best-fit Abundance \\
\hline
H$_2$O & $-6.26\pm0.14$ & $-6.34^{+0.63}_{-0.86}$  \\
CO     & $-5.44^{+0.23}_{-0.24}$ & $-5.44^{+0.65}_{-0.86}$  \\
CO$_2$ & $-9.63^{+0.99}_{-1.09}$ & $-10.47^{+1.29}_{-1.73}$  \\
FeH    & $-6.40^{+3.33}_{-3.40}$ & $-8.72^{+2.79}_{-0.15}$  \\
SiO    & $-5.08^{+0.18}_{-0.17}$ & $-3.98^{+2.41}_{-0.14}$  \\
TiO    & $-10.73^{+0.58}_{-0.48}$ & $-7.61^{+2.53}_{-0.34}$  \\
VO     & $-10.29^{+0.83}_{-0.79}$ & $-9.91^{+2.54}_{-0.45}$  \\
Fe     & $-6.45^{+3.37}_{-3.28}$ & $-3.81^{+2.75}_{-0.15}$   \\
Mg     & $-6.00\pm{3.35}$ & $-3.69^{+2.42}_{-0.14}$  \\
\hline
\multicolumn{3}{c}{\textbf{Inferred Bulk Composition}} \\
\hline
 & \uline{Chem. Eq. Retrieval} &  \uline{Chem. Eq. Fit} \\
{[M/H]}  & $0.40^{+0.43}_{-0.45}$ & $0.70^{+0.13}_{-0.11}$\\
C/O     & $0.01^{+0.01}_{-0.00}$& $0.0015^{+0.0007}_{-0.0005}$  \\
R/V     &$-0.18\pm0.49$ & $-1.33^{+0.16}_{-0.33}$\\
\hline
\end{tabular}
\caption{Top section: Retrieved $\log_{10}$ abundances of WASP-178b from \cite{lothringer:2025} using the isothermal free retrieval compared to our best-fit chemical equilibrium fit. Bottom section: Resulting bulk composition inferences the results of a chemical equilibrium retrieval (left) versus from fitting a chemical equilibrium solution to free retrieval results of the same observation with \exocomp{} (right).\label{tab:w178b}}
\end{table}

\subsubsection{WASP-178 b}

As another example, we fit a chemical equilibrium solution to the free retrieval results from the 0.2-5$\mu$m transmission spectrum of ultra-hot Jupiter WASP-178b \citep{lothringer:2025}. We choose this dataset as an example because 1) a corresponding chemical equilibrium retrieval was also performed in \cite{lothringer:2025}, 2) the variety of refractory and volatile species shows the utility of constraining Refr./Vol., and 3) we have ready access to the retrieval posteriors. To the latter point, we use WASP-178b as a demonstration of \exocomp{}'s ability to use input posterior samples instead of simple error bars, the latter of which will assume a Gaussian property that may not always apply. We choose the isothermal retrieval to compare to because the temperature at which to calculate the chemical equilibrium solution is given straightforwardly by the atmospheric temperature. We neglect the spurious constraint from Mg II, which we consider implausibly high and precise and which would otherwise dominated the refractory composition inference.

We use \exocomp{} to fit a chemical equilibrium solution to WASP-178b's isothermal free retrieval measurements with \texttt{PETRA}. Table~\ref{tab:w178b} lists the measurements from \exocomp{} and the \texttt{PETRA} retrieval side-by-side. We find [M/H] = $0.70^{+0.13}_{-0.11}$, C/O = $0.0015^{+0.0007}_{-0.0005}$, and Refr./Vol. = $-1.33^{+0.16}_{-0.33}$. These measurements broadly agree with the isothermal chemical equilibrium retrieval, which obtained [M/H] $\approx$ [O/H] = $0.4^{+0.43}_{-0.45}$, C/O = $0.01^{+0.01}_{-0.00}$, and Refr./Vol. $\approx$ [Si/O] = $-0.18\pm{0.49}$. Our estimated Refr./Vol. is somewhat lower than the retrieved [Si/O] ratio from the retrieval, likely due to the additional information provided by the TiO upper-limit, which will pull the R/V to lower values.

\section{The Library of Exoplanet Atmospheric Composition Measurements (LE\lowercase{x}AC\lowercase{o}M)}
\label{sec:table_analysis}
A homogenized collection of exoplanet composition measurements enables us to identify trends in planet composition with other planetary or stellar parameters. The identification of such trends allows us to test models of planet formation, which often make testable predictions for how a planet's composition varies with planet mass, disk mass, host star metallicity, etc. 

Here, we have used \exocomp{} to convert measurements of metallicity and C/O to elemental abundances based on the definition of metallicity and C/O and the solar abundance reference used in each study. Measurements are collected from a literature search of recent work, including using the ExoAtmospheres database\footnote{https://research.iac.es/proyecto/exoatmospheres/table.php} of the Instituto de Astrof\'isica de Canarias. Measurements were added if they included 1) observations of a companion to a main-sequence host star, 2) an explicitly stated measurement of metallicity and C/O, 3) has been accepted for publication, and 4) were obtained using data from either \jwst{} or an $\geq8$-meter-class ground-based observatory. The latter requirements ensured observations were of relatively high-quality data. In total, our sample consists of 66 measurements of 47 different planets. 16 unique planets are measured with direct spectroscopy, 15 with transmission spectroscopy, and 18 with eclipse spectroscopy.

The metallicity definition and retrieval used are both noted in the table, as these are necessary to convert metallicity into elemental abundances as discussed in Section~\ref{sec:bulk_to_elemental}; most models appear to have used \cite{asplund:2009} or \cite{asplund:2021} as the solar abundance definition, with \texttt{PETRA} sometimes using \citet{asplund:2005:abund} and PICASO using \cite{lodders:2010} for consistency with previous modeling in that framework. 

Some planets, like WASP-121b, have multiple measurements in the same geometry. Rather than choosing one measurement, we have included all applicable observations. When fitting for various trends below, we take the weighted average of these multiple measurements. This ensures that we do not weight the planets by the number of measurements available, while still incorporating information from each study agnostic to which might be closer to the ``true" measurement of a planet's composition. Each compositional measurement is also associated with other planetary and stellar properties, namely the planet's mass, equilibrium temperature, stellar metallicity and stellar mass, from the Planetary Systems Composite Data from the NASA Exoplanet Archive \citep{christiansen:2025}.

Table~\ref{tab:spec_stats} lists basic statistics, including the weighted average, standard deviation, and median for various sub-samples broken down by observing geometry and wavelength coverage. For the latter sub-sample, we include planets only observed from 1-3~$\mu$m with \jwst{}/NIRISS/SOSS, planets only observed from 3-5$\mu$m with either \jwst{}/NIRSpec/G395H, and then planets observed with high-resolution spectrographs. For transiting planets, we also filter out ultra-hot Jupiters to check if they bias the measurements in any way due to their extreme conditions (e.g., thermal inversions, molecular dissociation, etc.). Only a handful of explicit Refr./Vol. measurements currently exist in the literature, so here we focus solely on metallicity and C/O.

The weighted average can be heavily biased by a few precise measurements. An example of this is WASP-121b, whose precise high C/O measurement biases the transit spectroscopy average C/O, as well as the planets only observed from 3-5$\,\mu$m. In these cases, it is helpful to compare to the median values, which are not weighted and less biased by extreme measurements or outliers. We discuss the results of Table~\ref{tab:spec_stats} below in Sections~\ref{sec:met} and \ref{sec:co}.

\begin{deluxetable}{l l l l c c c l l c c c}
\tabletypesize{\small}
\tablecaption{Table of Exoplanet Atmospheric Composition Measurements (LE\lowercase{x}AC\lowercase{o}M) \label{tab:ExoCompTable}}
\tablehead{
\colhead{Planet} & 
\colhead{Reference} & 
\colhead{Status} & 
\colhead{Geometry} & 
\colhead{C/O} & 
\colhead{[M/H]} & 
\colhead{[M/H] Type} & 
\colhead{Solar Ref.} &
\colhead{Retrieval} &
\colhead{Mass ($M_J$)} &
\colhead{$T_{\rm eq}$ (K)} &
\colhead{[Fe/H]$_\star$}
}
\startdata
WASP-178 b & \cite{lothringer:2025} & Published & Transit & $0.01^{+0.01}_{-0.01}$ & $1.47^{+0.28}_{-1.10}$ & O/H & Stellar & \texttt{pRT} & $1.66^{+0.12}_{-0.12}$ & $2469^{+60}_{-60}$ & $0.21^{+0.16}_{-0.16}$ \\
HD 189733 b & \cite{fu:2024} & Published & Transit & $0.20^{+0.00}_{-0.20}$ & $0.60^{+0.10}_{-0.12}$ & (O+C)/H & Solar & \texttt{CHIMERA} & $1.13^{+0.08}_{-0.08}$ & $1193^{+11}_{-11}$ & $0.02^{+0.00}_{-0.00}$ \\
HD 209458 b & \cite{xue:2023} & Published & Transit & $0.08^{+0.09}_{-0.05}$ & $0.48^{+0.37}_{-0.18}$ & C/H & Solar & \texttt{PLATON} & $0.73^{+0.04}_{-0.04}$ & $1469^{+12}_{-12}$ & $0.01^{+0.00}_{-0.00}$ \\
HD 149026 b & B\cite{bean:2023} & Published & Eclipse & $0.84^{+0.03}_{-0.03}$ & $2.09^{+0.32}_{-0.35}$ & O/H & Solar & \texttt{PLATON} & $0.38^{+0.06}_{-0.06}$ & $1694^{+69}_{-37}$ & $0.32^{+0.00}_{-0.00}$ \\
HD 149026 b & \cite{gagnebin:2024} & Published & Eclipse & $0.67^{+0.06}_{-0.27}$ & $1.15^{+0.37}_{-0.27}$ & O/H & Solar & \texttt{PICASO} & $0.38^{+0.06}_{-0.06}$ & $1694^{+69}_{-37}$ & $0.32^{+0.00}_{-0.00}$ \\
\enddata
\tablecomments{This table is published in its entirety in the machine-readable format. A portion is shown here for guidance regarding its form and content. Full table references are: \cite{lothringer:2025},\cite{fu:2024},\cite{xue:2023},\cite{bean:2023},\cite{gagnebin:2024}, \cite{coulombe:2023}, \cite{brogi:2023}, \cite{line:2021}, \cite{reggiani:2022}, \cite{august:2023}, \cite{smith:2024}, \cite{pelletier:2025}, \cite{evans-soma:2025}, \cite{gapp:2025}, \cite{sikora:2025}, \cite{kanumalla:2024}, \cite{wiser:2025}, \cite{beatty:2024}, \cite{sing:2024}, \cite{welbanks:2024},\cite{mansfield:2024}, \cite{schlawin:2024}, \cite{xuan:2024}, \cite{molliere:2020}, \cite{gravity:2020}, \cite{nasedkin:2024b}, \cite{zhang:2023}, \cite{balmer:2025}, \cite{palma-bifani:2024}, \cite{hsu:2024}, \cite{gandhi:2025}, \cite{mayo:2025}, \cite{meech:2025}, \cite{kirk:2025}, \cite{liu:2025}, \cite{canas:2025}, \cite{ramkumar:2025}, \cite{brown-sevilla:2023}, \cite{gonzalez-picos:2025}, \cite{bazinet:2024}, \cite{verma:2025}, \cite{bachmann:2025}, \cite{zhang:2025}, \cite{finnerty:2025}, \cite{yang:2024}, \cite{lesjak:2023}, \cite{smith:2024a}, \cite{ahrer:2025}, \cite{ahrer:2025b}, \cite{lesjak:2025}, \cite{pelletier:2025b}, \cite{barat:2025}, \cite{voyer:2025}.}
\end{deluxetable}

\begin{deluxetable*}{lcccccc}
\tablecaption{Summary of atmospheric and physical properties of subsamples within the Library of Exoplanet Atmospheric Composition Measurements \label{tab:spec_stats}}
\tablehead{
\colhead{} & \colhead{C/O} & \colhead{[M/H]} & \colhead{O/H} & \colhead{C/H} & \colhead{Mass ($M_\mathrm{J}$)} & \colhead{$T_\mathrm{eq}$ (K)}
}
\startdata
\multicolumn{7}{c}{\textbf{Direct Spectroscopy (16 planets)}} \\
        Average & $0.611 \pm 0.090$ & $0.824 \pm 0.496$ & $9.138 \pm 0.545$ & $8.870 \pm 0.610$ & $13.02 \pm 7.21$ & $100.5 \pm 1.9$ \\
        Median  & 0.560 & 0.359 & 9.123 & 8.789 & 11.85 & 100.0 \\
\multicolumn{7}{c}{\textbf{Eclipse Spectroscopy (18 planets)}} \\
        Average & $0.773 \pm 0.187$ & $0.065 \pm 0.564$ & $9.070 \pm 0.815$ & $8.869 \pm 0.829$ & $2.70 \pm 2.73$ & $1678.9 \pm 773.3$ \\
        Median  & 0.705 & 0.394 & 9.094 & 8.788 & 1.72 & 1686.5 \\
\multicolumn{7}{c}{\textbf{Eclipse 
 No UHJ Spectroscopy (13 planets)}} \\
        Average & $0.729 \pm 0.204$ & $0.020 \pm 0.545$ & $9.115 \pm 0.829$ & $8.929 \pm 0.864$ & $2.36 \pm 2.32$ & $1360.5 \pm 672.5$ \\
        Median  & 0.720 & 0.550 & 9.260 & 8.941 & 1.67 & 1411.0 \\
\multicolumn{7}{c}{\textbf{Transit Spectroscopy (15 planets)}} \\
        Average & $0.705 \pm 0.427$ & $0.957 \pm 0.710$ & $9.546 \pm 0.727$ & $9.033 \pm 0.710$ & $1.47 \pm 2.12$ & $1518.0 \pm 613.5$ \\
        Median  & 0.350 & 0.930 & 9.593 & 8.786 & 0.78 & 1543.0 \\
\multicolumn{7}{c}{\textbf{Transit 
 No UHJ Spectroscopy (12 planets)}} \\
        Average & $0.294 \pm 0.125$ & $0.885 \pm 0.813$ & $9.445 \pm 0.777$ & $9.055 \pm 0.713$ & $1.32 \pm 2.30$ & $1299.3 \pm 478.9$ \\
        Median  & 0.350 & 0.584 & 9.348 & 8.761 & 0.69 & 1371.0 \\
\multicolumn{7}{c}{\textbf{UHJs Spectroscopy (8 planets)}} \\
        Average & $0.728 \pm 0.419$ & $0.860 \pm 0.523$ & $9.327 \pm 0.779$ & $8.800 \pm 0.709$ & $3.02 \pm 2.88$ & $2463.7 \pm 155.8$ \\
        Median  & 0.497 & 0.895 & 9.646 & 8.737 & 1.82 & 2459.0 \\
\multicolumn{7}{c}{\textbf{1-3um Spectroscopy (2 planets)}} \\
        Average & $0.815 \pm 0.032$ & $0.533 \pm 0.606$ & $9.281 \pm 0.579$ & $9.127 \pm 0.646$ & $5.68 \pm 4.52$ & $2476.5 \pm 27.5$ \\
        Median  & 0.710 & 0.626 & 9.281 & 9.127 & 5.68 & 2476.5 \\
\multicolumn{7}{c}{\textbf{3-5um Spectroscopy (18 planets)}} \\
        Average & $0.943 \pm 0.132$ & $0.762 \pm 0.828$ & $9.473 \pm 0.809$ & $9.119 \pm 0.856$ & $1.05 \pm 0.93$ & $1451.2 \pm 563.6$ \\
        Median  & 0.500 & 0.641 & 9.477 & 8.973 & 0.95 & 1585.5 \\
\multicolumn{7}{c}{\textbf{Hi-Res Spectroscopy (16 planets)}} \\
        Average & $0.553 \pm 0.063$ & $0.057 \pm 0.534$ & $8.710 \pm 0.749$ & $8.502 \pm 0.711$ & $4.98 \pm 6.27$ & $1739.4 \pm 763.1$ \\
        Median  & 0.635 & -0.190 & 8.430 & 8.341 & 1.89 & 1691.0 \\
\multicolumn{7}{c}{\textbf{T-Dwarf Sample (50 brown dwarfs, \citealt{zalesky:2022})}} \\
    Average & $0.77 \pm 0.20$ & $-0.20 \pm 0.24$ & -- & -- & -- & $787.6 \pm 94.8$ \\
    Median  & 0.775 & $-0.22$ & -- & -- & -- & 775.7 \\
\multicolumn{7}{c}{\textbf{Solar Neighborhood (355 FGKM stars, \citealt{hinkel:2014})}} \\
    Average & $0.326 \pm 0.143$ & $-0.084 \pm 0.263$ & $8.732 \pm 0.213$ & $8.293 \pm 0.246$ & $1.10 \pm 0.21$ $M_\odot$ & $5866.0 \pm 577.7$ \\
    Median  & 0.468 & $-0.060$ & 8.720 & 8.390 & 1.07 $M_\odot$ & 5881.0 \\
\enddata
\tablecomments{Averages are inverse-variance weighted. Standard deviations are included in the average values to represent the scatter seen in the population, not necessarily as an actual uncertainty on the value.}
\end{deluxetable*}

\subsection{Metallicity (O/H)}\label{sec:met}

Some early \hst{} and \spit{} measurements were interpreted to indicate a general sub-solar-to-solar metallicity trend for giant planets \citep[e.g.,][]{madhusudhan:2014b,welbanks:2019a,barstow:2017}. These studies relied heavily on the single resolved H$_2$O feature at 1.4$\,\mu$m with \hst{}/WFC3/G141. On the other hand, \cite{walsh:2025} showed that such a handful of hot Jupiter measurements implied a general metal-enrichment. With the expanded wavelength coverage and larger telescope apertures enabled by current facilities, we assess here whether a general trend holds up across the gas-giant population.

\begin{figure*}
    \centering
    \includegraphics[width=0.88\linewidth]{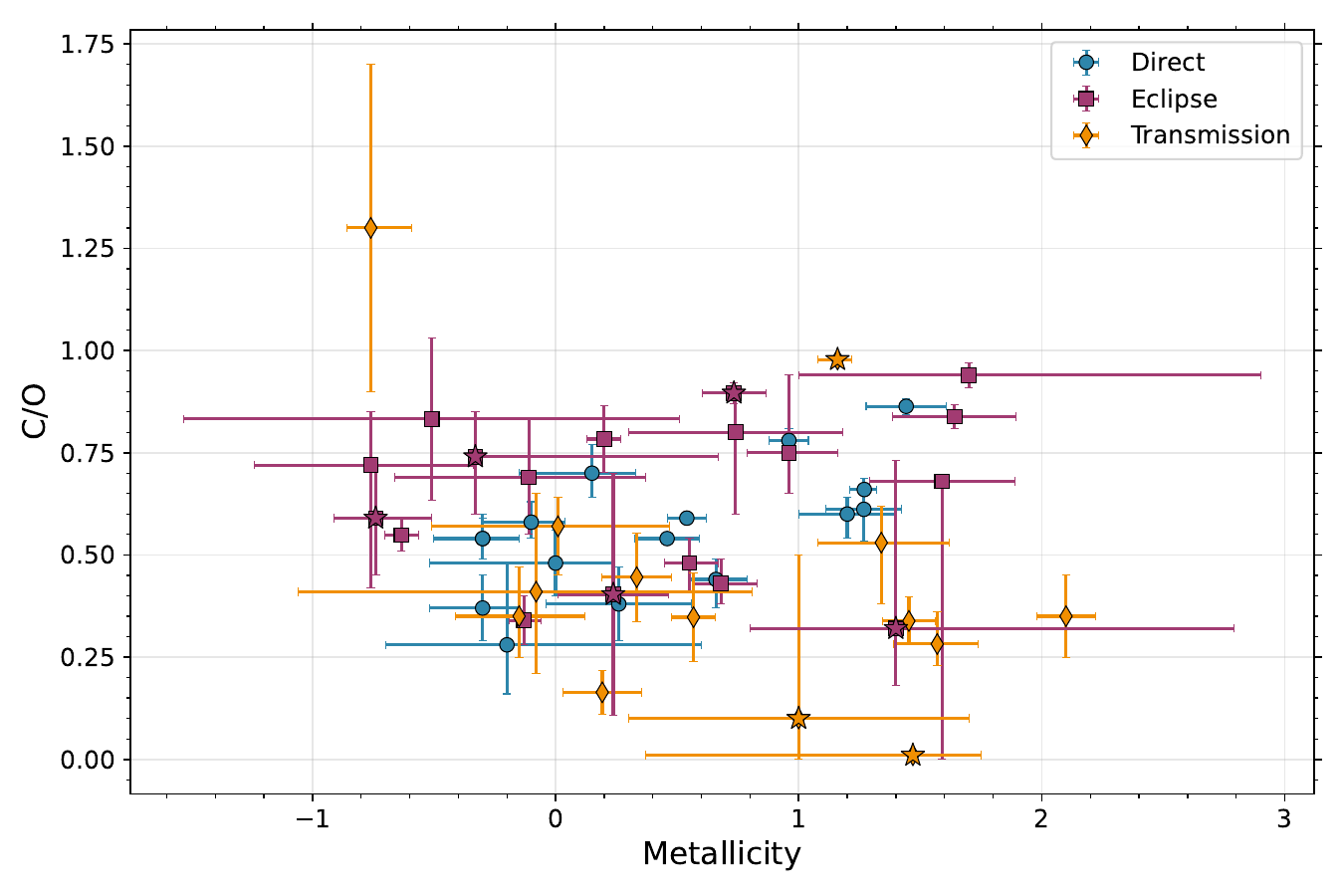}
    \caption{C/O ratio as a function of metallicity [M/H] as given in Table~\ref{tab:ExoCompTable}. Planets observed with direct spectroscopy are shown as blue circles, planets observed with eclipse spectroscopy as magenta squares, and planets observed with transit spectroscopy as gold diamonds. Ultra-hot Jupiters are labeled as stars.}
    \label{fig:met_co}
\end{figure*}

Figure~\ref{fig:met_co} shows the sample's metallicity versus C/O. Across the whole sample, metallicity is found to vary from about $0.1-100\times$ solar. Some of the highest metallicity measurements are from low-mass planets, which we discuss in the context of the mass-metallicity trend in Section~\ref{sec:mass_met}. We also discuss the surprisingly large scatter in metallicity for planets around $1\,M_{\mathrm{Jup}}$ in Section~\ref{sec:mass_met}. 

\begin{figure*}
    \centering
    \includegraphics[width=0.88\linewidth]{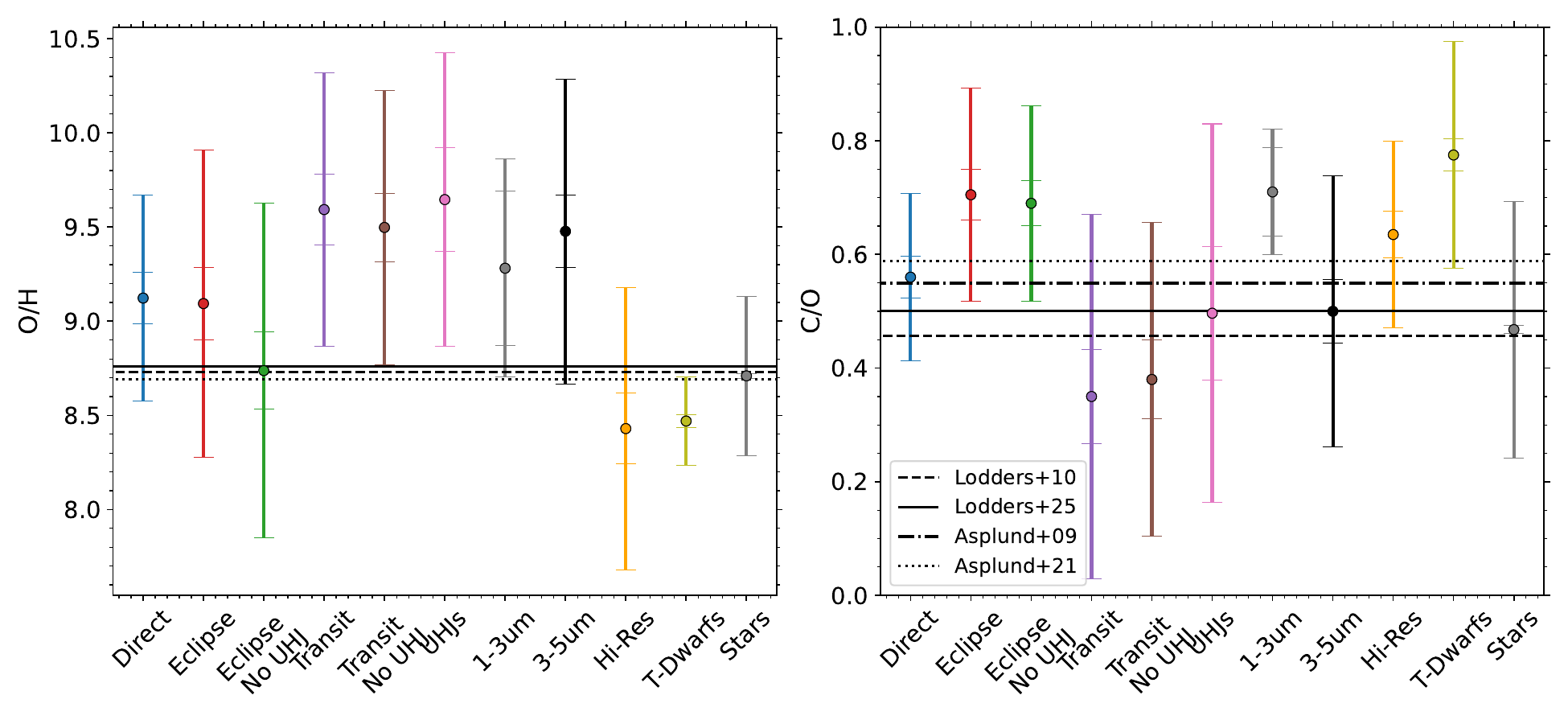}
    \caption{Median values of the O/H (left, $\log_{10}$ parts per trillion) and C/O (right) shown for different sub-samples. The outer-most errorbar shows the standard deviation of measurements from that sample, while the inner errorbar represents the standard deviation of the mean for reference. Also plotted as horizontal black lines are O/H and C/O values from different stellar abundance definitions.}
    \label{fig:met_co_geo}
\end{figure*}

Figure~\ref{fig:met_co_geo} shows the median of the O/H abundance, calculated from the metallicity using \exocomp{}. Also shown are the standard deviation of the sample, illustrating the wide range of measured values, and the standard deviation of the mean. We note that the standard deviation of the mean is useful statistically when measurements in a sample come from the same distribution, which likely does not hold for the sub-samples of planets that stretch across over an order of magnitude in mass, so we include it here only as a description of the scatter. 

Each sample of planets in Figure~\ref{fig:met_co_geo} shows an enrichment in heavy elements above solar values. The enrichment holds when compared to stars in the local solar neighborhood from the Hypatia catalog \citep{hinkel:2014} and when compared to T-dwarfs with measured metallicities from atmosphere retrievals of near infrared spectra \citep{zalesky:2022}. The former is perhaps unsurprising given the large number of relatively low-metallicity stars and the group of high-metallicity exoplanet measurements. The latter, however, is important because it uses the same techniques used to measure the exoplanet compositions so any major systematic bias from our measurement techniques could be present in that sample as well. 

We perform two-sample Kolmogorov-Smirnov (K-S) tests to assess whether a given sample differs from another in a statistically meaningful way. A two-sample K-S test is a non-parametric method that compares the largest distance (i.e., the test statistic) between two samples' empirical cumulative distributions. We calculate the p-value from the test statistic using the asymptotic distribution method as implemented in \texttt{scipy.stats.kstest}.

An unweighted K-S test shows no statistically significant difference in the metallicity between subsets of the total exoplanet sample. Each population shows a similar spread in metallicity, with averages that are slightly super-solar. However, the full exoplanet sample does appear to be statistically different then the stellar sample with a p-value of $6.3\times10^{-19}$, with exoplanets showing a general metal-enrichment compared to the stellar population, consistent with the empirical trends from \cite{fu:2025}.

We also performed a inverse-variance weighted K-S test following methods from \cite{press:2002}, resulting in a somewhat different story. With the weighted sample, the sample of planets observed with direct and transit spectroscopy were both found to be statistically different from the sample of planets observed with eclipse spectroscopy, with p-values of 0.0018 and 0.0023, respectively. This difference may be due to the fact that planets observed with eclipse spectroscopy can reach lower metallicities than those with direct or transit spectroscopy. Like with the unweighted K-S test, the weighted test indicated a very significant difference between the full exoplanet population and the stellar population, with a p-value of $8.2\times10^{-28}$.

The difference between interpretations for the unweighted versus weighted K-S tests is somewhat surprisingly, but could be attributed to the very high weights of a few precise measurements in each population. Increasing the sample size for each observing geometry will help in distinguishing statistical differences between the populations.

\subsubsection{Metallicity-Temperature}\label{sec:met_temp}

\begin{figure}
    \centering
    \includegraphics[width=1\linewidth]{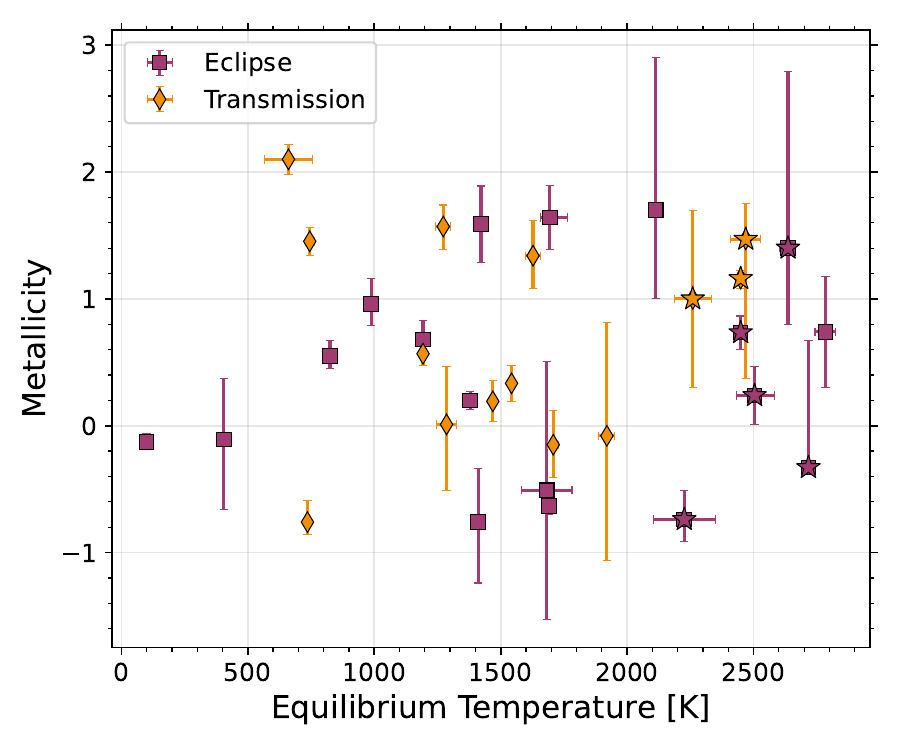}
    \caption{Metallicity as a function of equilibrium temperature for planets in the eclipse (magenta squares) and transmission (gold diamonds) spectroscopy samples. Ultra-hot Jupiters are labeled as stars.}
    \label{fig:temp_met}
\end{figure}

Correlations with temperature could indicate biases in the retrievals due to temperature-dependent processes. Of particular importance is H$_2$O thermal dissociation in ultra-hot Jupiters \citep{kitzmann:2018,parmentier:2018,lothringer:2018b}, which will deplete one of the main metallicity indicators in hot Jupiters. Figure~\ref{fig:temp_met} shows the relationship between equilibrium temperature and metallicity for the transit and eclipse spectroscopy sample. As with C/O ratio, we do not include the direct spectroscopy sample here as we would need to compare to the measured planetary effective temperature, which is model dependent and was not collected here.

Here, though, we find no significant correlation between temperature and metallicity, using both weighted and unweighted correlation coefficients. Unweighted correlation coefficients were below 0.14 for both the emission and transmission spectroscopy samples.

\subsubsection{Mass-Metallicity}\label{sec:mass_met}

\begin{figure*}[ht!]
    \centering
    \includegraphics[width=0.85\linewidth]{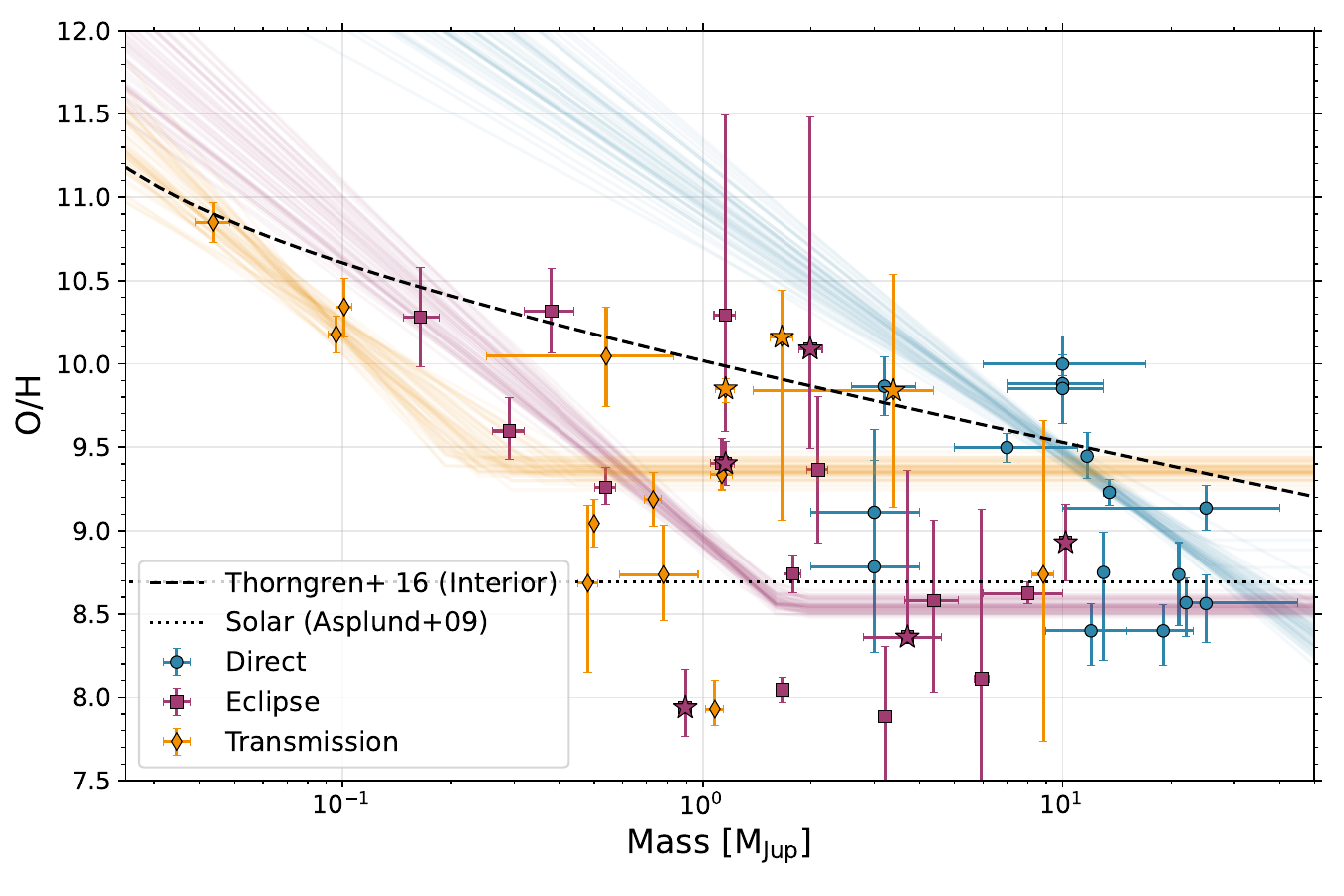}
    \includegraphics[width=0.85\linewidth]{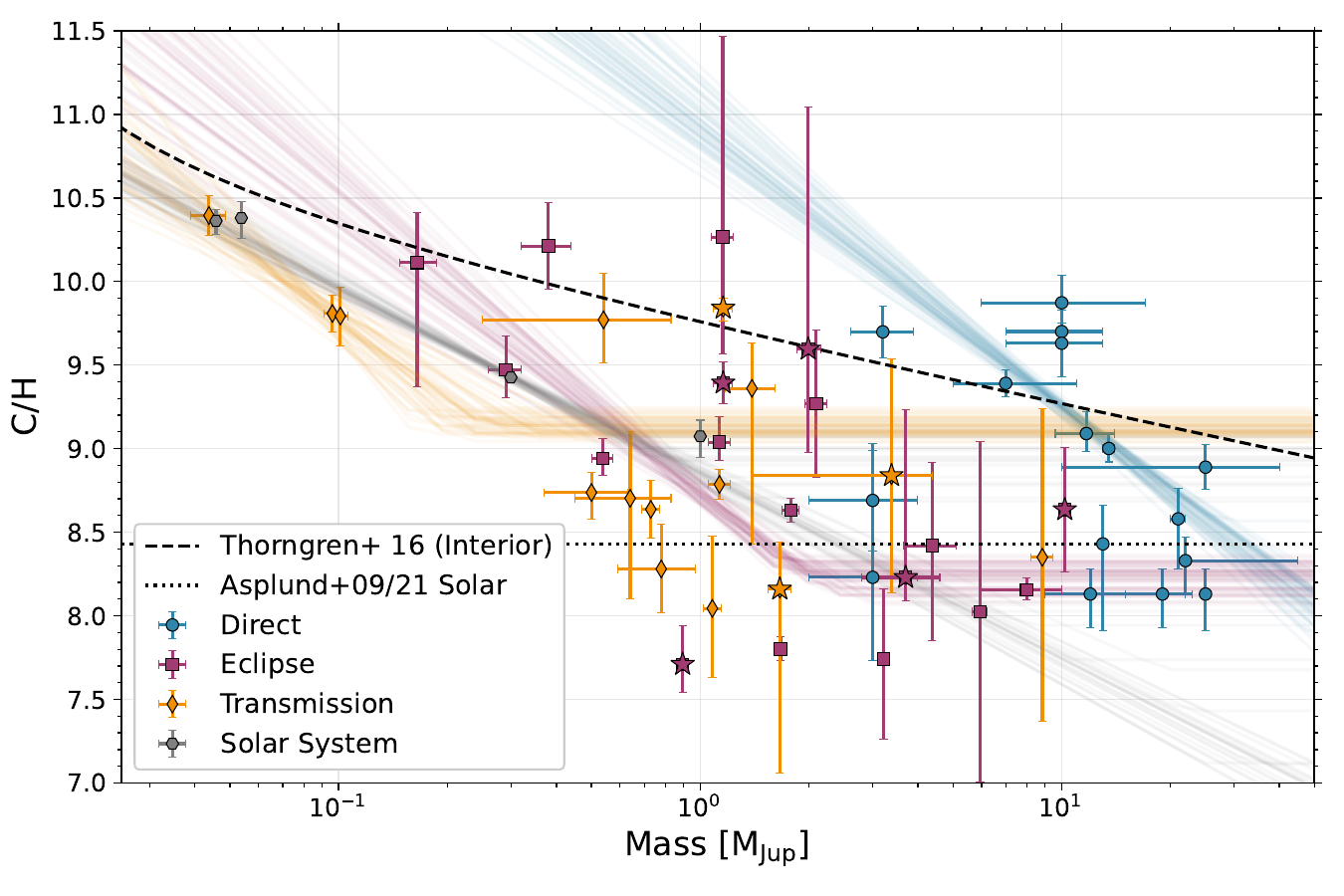}

    \caption{O/H (top) and C/H (bottom) in $\log_{10}$ parts per trillion as a function of mass. Planets observed with direct spectroscopy are shown as blue circles, planets observed with eclipse spectroscopy as magenta squares, and planets observed with transit spectroscopy as gold diamonds. The Solar System gas planets are shown in grey and ultra-hot Jupiters are labeled as stars. 50 samples from the fit using Eq.~\ref{eq:mass-metallicity} are shown for reference. The dotted line shows the bulk metallicity versus mass trend inferred from interior models \citep{thorngren:2016} converted to number density via Eq.~2 of \cite{thorngren:2019a}.}
    \label{fig:mass_metallicity_fit}
\end{figure*}

The relationship between a planet's mass and its metallicity (whether global/planet-wide/bulk or atmospheric) has been hypothesized to be a tell-tale signal of formation via core-accretion \citep{pollack:1996}. Lower-mass planets are thought to have a higher proportion of heavy elements because their high-metallicity core makes up a greater proportion of their total mass, while also being able to efficiently accrete planetesimals to enrich the atmosphere \citep[e.g.,][]{miller:2011,fortney:2013,mordasini:2014,mordasini:2016,swain:2024}. High-mass planets will tend to accrete most of their mass through gas accretion, including a large proportion of hydrogen and helium, resulting in lower metallicities.

The mass-metallicity anti-correlation has been shown to hold for measurements of the atmospheric metallicity of Solar System giant planets, as well as early hot Jupiter measurements \citep[e.g.,][]{kreidberg:2014b,wakeford:2018,wakeford:2020b,swain:2024}, though some studies found a lack of a relationship between $0.1-1\,M_{\mathrm{Jup}}$ \cite{pinhas:2019}. \cite{sun:2024} used an updated catalog of \hst{} measurements, finding a lack of a statistically significant trend, in part because of confounding measurements of low-mass, low-metallicity planets like HAT-P-11b \citep{chachan:2019}. Using a wider array of metallicity tracers, including Na and K, \cite{welbanks:2019a} found a significant mass-metallicity relationship, but with lower O/H abundances than in the Solar System.

Measurements of planetary masses and radii reflect a mass-metallicity trend \citep{thorngren:2016}, where planets with higher bulk metallicities will have a smaller radius for the same mass compared to a planet with lower bulk metallicity. Put simply, this is due to the fact that materials with more heavy elements will tend to be denser. Notably, these techniques measure not just the metallicity of the atmosphere/envelope, but are sensitive to the entire planet's metal content, including the interior layers. The degree to which the atmospheric metallicity reflects the bulk metallicity is an open area of research, as mixing in giant planet interiors is not well constrained \citep[e.g.,][]{guillot:1994b,leconte:2012,vazan:2015,wahl:2017,vazan:2018,stevenson:2020,howard:2023,muller:2024}. Differences between the metallicity of the atmosphere and interior can provide insight into such mixing \citep{thorngren:2019a}.

Here, we use Table~\ref{tab:ExoCompTable} to constrain the presence and properties of the mass-metallicity trend as measured by transmissions, emission, and direct spectroscopy. We fit a linear trend between mass and metallicity, with a freely-fit cut-off mass, above which the metallicity is constant with mass:

\begin{equation}\label{eq:mass-metallicity}
M/H(M_p) = \left\{
  \begin{array}{ll}
    s \times \log_{10}(M_p) + b & \text{if } M_p  < M_c \\
    s \times \log_{10}(M_c) + b & \text{if } M_p  \geq M_c
  \end{array}
\right.
\end{equation}

\noindent where $M/H$ is the metallicity (either O/H, C/H, or some other definition), $M_p$ is the planet's mass, $M_c$ is the cut-off mass, $s$ is the slope of the linear trend, and $b$ is the intercept (effectively the metallicity at $1 \,M_{\mathrm{Jup}}$). We fit O/H and C/H as $\log_{10}$ parts per trillion, in line with the solar abundance convention.

We fit the sample independently with respect to observing geometry. We include an similarly independent fit to the measurements from Jupiter, Saturn, Uranus, and Neptune \citep[e.g.,][respectively]{wong:2004,fletcher:2009b,sromovsky:2011,karkoschka:2011} taken from \cite{kreidberg:2014b}. When comparing to the Solar System, it is important to note that the metallicities inferred are all based on measurements of CH$_4$, while for the exoplanets, H$_2$O, CO$_2$, and CO are usually the measured species. Because of this, we fit the mass-metallicity trend with both O/H and C/H, converted from the metallicity in Table~\ref{tab:ExoCompTable} with \exocomp{}, accounting for the fact that metallicity is defined differently in various retrieval codes. We used the \texttt{Dynesty} nested sampler \citep{koposov:2024}, which allows us to measure the Bayesian evidence to compare with a model where there was no relationship between mass and metallicity (i.e., a metallicity that is constant with mass).

In Figure~\ref{fig:mass_metallicity_fit}, we show the sample metallicity, as represented by the O/H and C/H ratio, as a function of mass. Also plotted are 50 random samples from the nested sampling fitting for each observing geometry. While significant scatter exists in the mass-metallicity parameter space around $1\,M_{\mathrm{Jup}}$ as mentioned below, the handful of measured low-mass eclipse and transmission spectroscopy targets clearly have elevated metallicities relative to the higher mass planets. We find statistically significant mass-metallicity trends relative to a constant-with-mass metallicity fit, with differences in the log-Bayesian evidence (i.e., the Bayes factor) often greater than 50.

Surprisingly, there appears to be a large scatter in metallicity for planets at and above $1\,M_{\mathrm{Jup}}$. This scatter appears to exist in both transiting and non-transiting planets, both with standard deviations of about half a dex or greater in [M/H] (see Table~\ref{tab:spec_stats}). While the scatter may be a sign of retrievals struggling to measure accurate metallicities across the Jovian exoplanet population, this could indicate a somewhat unexpected astrophysical diversity of Jovian planet compositions. 

Using population synthesis planet formation modeling, \cite{mordasini:2014} showed an expected scatter of between $1-50\times$ solar metallicity with higher mass planets reaching metallicities at or below 1$\times$ solar through gas-dominated accretion and lower-mass planets reaching metallicities up to 100$\times$ solar. However, in the currently observed sample, there are several high-mass planets with metallicity at or above $10\times$ solar metallicity, as well as a population of Jovian-mass planets below $1\times$ solar metallicity. We show below that the planetary metallicity does not tend to correlate with equilibrium temperature (Sec~\ref{sec:met_temp}) or stellar mass and metallicity (Sec~\ref{sec:stellar}), leaving the door open to possibilities beyond the scope of this study.

Unexpectedly, we also see a mass-metallicity trend with the sample of planets measured with direct spectroscopy. In general, these planets are much higher in mass than the planets measured with transmission or eclipse spectroscopy, as indicated in Table~\ref{tab:spec_stats}, approaching the upper-limit for which atmospheric enrichment via planetesimal accretion is thought to be effective \citep{mordasini:2016}. As such, one would expect these planets to be higher than any cut-off mass for elevated enrichment during the planet formation process; any planet would need to be enriched by more rocky or icy material than is thought to be available to achieve a high enough heavy metal fraction to overcome the large amount of gas accreted by such planets. We note that the mass-metallicity trend for direct spectroscopy planets appears to mostly be driven by the VLT/SPHERE and GRAVITY measurement of elevated metallicity in AF Lep ($\sim10\times$ solar metallicity) by \cite{zhang:2023} and \cite{balmer:2025}, which are the highest precision measurements at the low-mass end of the sample. We suggest that the linear mass-metallicity trend is a poor representation of the scatter in metallicity for planets observed with direct spectroscopy. More observations of this and other low-mass targets with direct spectroscopy may better constrain the population-level behavior and scatter.

\begin{figure*}
    \centering
    \includegraphics[width=1\linewidth]{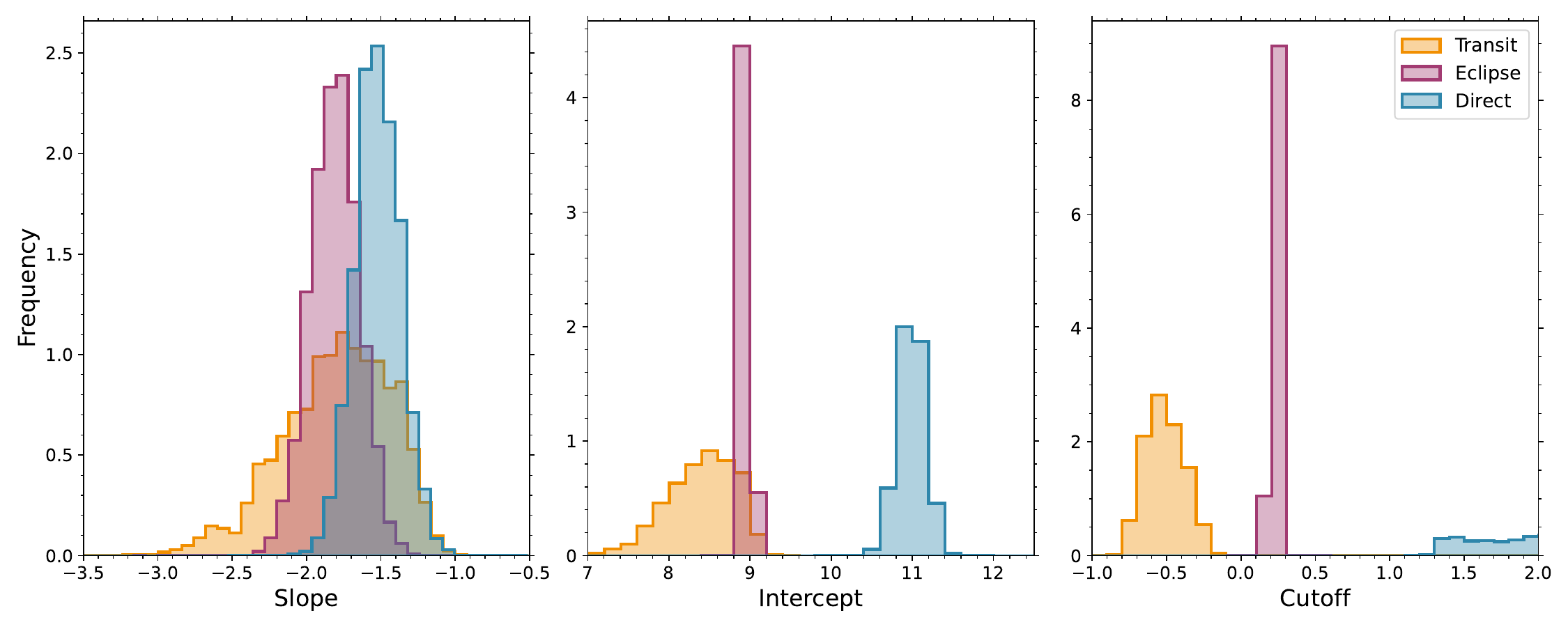}
    \includegraphics[width=1\linewidth]{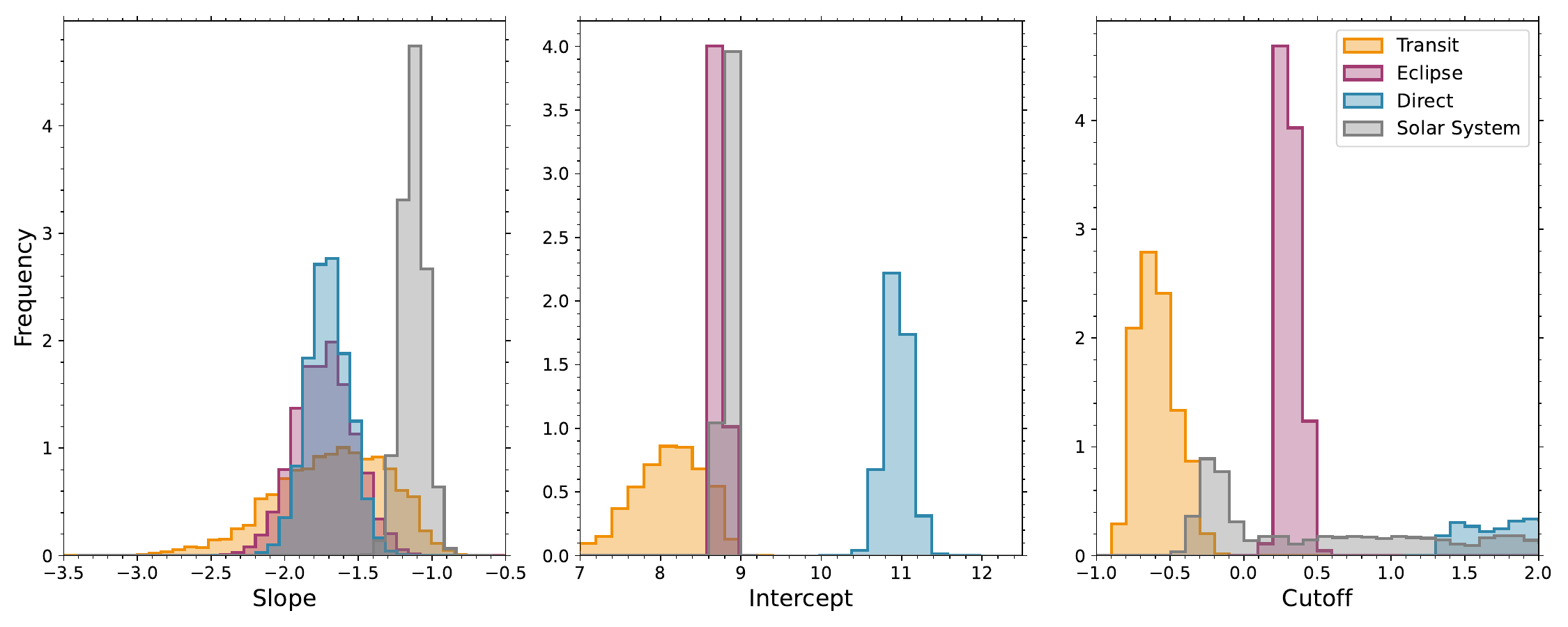}
    \caption{Top: Mass-metallicity trend parameter constraints using Eq.~\ref{eq:mass-metallicity} for the different observing geometries using O/H as the metallicity indicator. Bottom: Same as top, but using C/H as the metallicity indicator.}
    \label{fig:mass_metallicity_posterior}
\end{figure*}

Figure~\ref{fig:mass_metallicity_posterior} and Table~\ref{tab:mass_metallicity} shows the measured values from the fits. All measurements agree whether we use O/H or C/H as our metallicity indicator. In Table~\ref{tab:mass_metallicity} we also show a fit where we used the metallicity as given in Table~\ref{tab:ExoCompTable}, rather than converting to O/H or C/H, as a test. We discuss the values for the slope, intercept, and cut-off mass individually below.

\begin{table*}
\centering
\begin{tabular}{lcccc}
\multicolumn{5}{c}{\textbf{Metallicity as Given}} \\
\hline
\textbf{Parameter} & \textbf{Transiting} & \textbf{Eclipse} & \textbf{Direct} & \textbf{Solar System} \\
\hline
Slope & $-1.78\pm0.35$ & $-1.75\pm0.16$ & $-1.71\pm0.15$ & $-1.13\pm0.09$ \\
Intercept (M/H) & $-0.32\pm0.40$ & $0.28\pm0.04$ & $2.51\pm0.16$ & $0.41\pm0.07$ \\
Mass cut-off (log10) & $-0.51\pm0.13$ & $0.22\pm0.02$ & $3.18\pm1.07$ & $1.83\pm1.71$ \\
$\ln E$ & $-117.96 \pm 0.2$ & $-153.57 \pm 0.2$ & $-146.24 \pm 0.2$ & $-11.85 \pm 0.2$ \\
\hline 
\multicolumn{5}{c}{\textbf{Flat-Line}} \\
\hline
\textbf{Parameter} & 	\textbf{Transiting} & \textbf{Eclipse} & \textbf{Direct} & \textbf{Solar System} \\
\hline 
Intercept (M/H) & $0.94\pm0.04$ & $0.11\pm0.03$ & $0.71\pm0.03$ & $1.04\pm0.02$ \\
$\ln E$ & $-205.81 \pm 0.2$ & $-221.26 \pm 0.2$ & $-209.34 \pm 0.2$ & $-105.52 \pm 0.2$ \\
\hline
\\[-0.5ex]
\\[-0.5ex]
\multicolumn{5}{c}{\textbf{O/H}} \\
\hline
\textbf{Parameter} & \textbf{Transiting} & \textbf{Eclipse} & \textbf{Direct} & \textbf{Solar System} \\
\hline
Slope & $-1.80\pm0.36$ & $-1.81\pm0.16$ & $-1.52\pm0.16$ & -- \\
Intercept (M/H) & $8.40\pm0.41$ & $8.95\pm0.04$ & $10.98\pm0.17$ & -- \\
Mass cut-off (log10) & $-0.51\pm0.12$ & $0.22\pm0.02$ & $3.15\pm1.06$ & -- \\
$\ln E$ & $-111.21 \pm 0.2$ & $-143.03 \pm 0.2$ & $-125.21 \pm 0.2$ & -- \\
\hline
\multicolumn{5}{c}{\textbf{Flat-Line}} \\
\hline
\textbf{Parameter} & 	\textbf{Transiting} & \textbf{Eclipse} & \textbf{Direct} & \textbf{Solar System} \\
\hline 
Intercept (M/H) & $9.69\pm0.04$ & $8.81\pm0.03$ & $9.37\pm0.03$ & --\\
$\ln E$ & $-198.82 \pm 0.2$ & $-211.85 \pm 0.2$ & $-167.27 \pm 0.2$ & $-105.59 \pm 0.2$ \\
\hline
\\[-0.5ex]
\\[-0.5ex]
\multicolumn{5}{c}{\textbf{C/H}} \\
\hline
\textbf{Parameter} & \textbf{Transiting} & \textbf{Eclipse} & \textbf{Direct} & \textbf{Solar System} \\
\hline
Slope & $-1.68\pm0.38$ & $-1.73\pm0.20$ & $-1.71\pm0.14$ & $-1.13\pm0.09$ \\
Intercept (M/H) & $8.10\pm0.44$ & $8.74\pm0.04$ & $10.95\pm0.15$ & $8.84\pm0.07$ \\
Mass cut-off (log10) & $-0.59\pm0.13$ & $0.30\pm0.08$ & $3.15\pm1.06$ & $1.84\pm1.71$ \\
$\ln E$ & $-126.77 \pm 0.2$ & $-161.91 \pm 0.2$ & $-146.22 \pm 0.2$ & $-11.92 \pm 0.2$ \\
\hline 
\multicolumn{5}{c}{\textbf{Flat-Line}} \\
\hline
\textbf{Parameter} & \textbf{Transiting} & \textbf{Eclipse} & \textbf{Direct} & \textbf{Solar System} \\
\hline 
Intercept (M/H) & $9.39\pm0.04$ & $8.52\pm0.03$ & $9.15\pm0.03$ & $9.47\pm0.02$ \\
$\ln E$ & $-180.26 \pm 0.2$ & $-224.50 \pm 0.2$ & $-208.56 \pm 0.2$ & $-105.42 \pm 0.2$ \\
\end{tabular}
\caption{Parameter estimates from fits to the mass-metallicity trend using Eq.~\ref{eq:mass-metallicity} for different definitions of metallicity. \label{tab:mass_metallicity}
}
\end{table*}

\paragraph{Mass-Metallicity Slope}

All samples show a negative slope between $-1$ and $-2$ dex in $M/H$ per dex in mass, consistent with the expected mass-metallicity trend from planet formation \citep[e.g.,][]{mordasini:2014}. Of note is the agreement in the mass-metallicity slope between planets measured in transmission versus eclipse spectroscopy, which agree to well within 1-$\sigma$. This suggests that the mass-metallicity trend we measure here is robustly astrophysical as it is replicated in two samples that are measured in with two different observational setups and retrieved with fundamentally different modeling.

Interestingly, the Solar System mass-metallicity trend shows the shallowest slope at $-1.127 \pm 0.08$ dex in $M/H$ per dex in mass. While limited to a sample of only four planets, this may suggest that the mass-metallicity trend may vary from system-to-system. \cite{helled:2014} highlighted the sensitivity of planets like Uranus and Neptune to the conditions of their formation, expecting a diversity of compositions for intermediate-mass exoplanets. 

With that said, the Solar System planets do seem well-approximated by the fit to the transmission spectroscopy sample. It is only Saturn's metallicity that diverges slightly from the transmission spectroscopy mass-metallicity trend. Where the transmission spectroscopy sample prefers a mass cut-off around Saturn's mass, the Solar System sample is well-fit by either a higher mass cut-off or no mass cut-off at all. Interestingly, the eclipse spectroscopy sample does not fit the Solar System mass-metallicity trend well, finding much higher metallicities for Neptune-mass exoplanets. This is perhaps reflective of the challenge of measuring the composition low- and intermediate-mass exoplanets with eclipse spectroscopy, even in the \jwst{} era \citep[e.g.,][]{mukherjee:2025}. Future observations of the metallicity for planet's in multi-planet systems could shed light on the degree of intrinsic scattering in the mass-metallicity trend from system to system.

We can also compare to the mass-metallicity trend from analyses of the relationship between planet mass and radius as interpreted through interior models. Higher metallicity planets will tend to have a larger mass for a given radius (i.e., a higher density), enabling a constraint on bulk (not just atmospheric) metallicity. Using this technique, \cite{thorngren:2016} found a slope of $-0.45 \pm 0.09$ between planet mass and metal enrichment (in terms of mass fraction). This is significantly shallower than the slope found by our comparison to the atmospheric metallicity as a function of mass. We also note that planet formation population synthesis models \citep{mordasini:2014} also find a similarly shallow slope as \cite{thorngren:2016} at -0.68. 

If we assume that the interior has a greater metallicity than the envelope because of differentiation (i.e., a heavy-element core), then a shallower slope from interior models compared to atmospheric constraints would imply that the atmosphere becomes more metal rich with decreasing mass \emph{faster} than the interior does, which could also point to lower mass planets being more well-mixed. We discuss this in more detail below.

\paragraph{Mass-Metallicity Intercept}

The intercept of the trend, $b$, represents the metallicity where $\log_{10}(M_p) = 0$, which in this case means at $1 \,M_{\mathrm{Jup}}$. For eclipse spectroscopy, this corresponds to approximately $2\times$ solar metallicity. For transmission spectroscopy, this would correspond to $0.5\times$ solar metallicity if the trend continued, but the mass cut-off fixes the actual metallicity at $1\,M_{\mathrm{Jup}}$ to be about 4.9$\times$ solar. Both the values from eclipse and transmission spectroscopy are close to the estimate of Jupiter's metallicity of between 3.3 and 5.5$ \times$ solar \citep{wong:2004}. The agreement is somewhat surprising given the amount of scatter in the metallicity measurements around $1\,M_{\mathrm{Jup}}$, which vary from $10\times$ sub-solar to $10\times$ super-solar, yet appear to average out to agree with the Solar System value.

Another interesting comparison is again with the constraints on bulk metallicity from interior modeling by \cite{thorngren:2016}. The bulk metallicity at $1\,M_{\mathrm{Jup}}$ from the interior modeling (converted from mass-fraction to number-fraction \citep{thorngren:2019a}) is several times higher than the atmospheric metallicity of Jupiter, consistent with the idea that Jupiter's interior is more metal enriched than its envelope \citep[e.g.,][]{hubbard:1989,wahl:2017}. If the enrichment trend from \cite{thorngren:2016} represents the true bulk metallicity of these exoplanets, then their distance below the line would indicate increasing levels of differentiation. Planets closer to the line would represent scenarios where the planets are well-mixed. Even the high-measured metallicities of high-mass planets, like in the HR 8799 system, are not that far off from the population-wide bulk enrichment trend -- with the caveat that these specific planets must then be well-mixed. 

The fact that lower-mass planets appear more well-mixed (i.e., the atmospheric metallicity is closer to the bulk metallicity) has important implications for planet formation and evolution. One explanation would be that lower-mass planets have a lower core mass fraction. If this low core mass fraction is primordial, then such planets must have formed with a relatively low-mass core before accreting a gaseous envelope, perhaps at a reduced accretion rate, explaining their present-day low-mass. Alternatively, this low core mass fraction could be due to mixing in the interior, effectively eroding the primordial core, like in \textit{Juno} gravity measurements of Jupiter \citep{wahl:2017}.

The level of mixing a planet undergoes will shape a planet's luminosity and cooling over time \citep{nettelmann:2013,vazan:2020b}, and determine the type of atmosphere we see at present day. A trend of increasing mixing with decreasing planet mass is also tentatively consistent with the idea that the interior of planets like Uranus and Neptune may be more of a rock-ice mix, rather than a fully-layered model \citep{vazan:2022}, with strong compositional gradients in more massive planets preventing efficient mixing \citep[e.g.,][]{leconte:2012,moll:2017,vazan:2018}. With the spread in metallicity for planets around $1\,M_{\mathrm{Jup}}$ as noted in Sec.~\ref{sec:met} and \ref{sec:mass_met}, it opens up the intriguing possibility that this scatter could be evidence for a variance in the amount of mixing with the interior. Future modeling of planetary interiors, planet formation, and planet evolution can investigate the relative importance of mixing versus enrichment during formation in replicating the distribution of giant exoplanet atmospheric metallicities.

We discount the mass-metallicity trends from direct spectroscopy as unphysical, as it would imply a metallicity at $1\,M_{\mathrm{Jup}}$ of about $1000\times$ solar. Again, this is not because of the measurements themselves, but rather that it is likely that a linear mass-metallicity trend is a poor representation of the scatter in metallicity for planets observed with direct spectroscopy.

\paragraph{Mass-Metallicity Cut-Off Mass}

In the context of planet formation models, we define ``cut-off mass'' as the threshold above which a planet's metallicity becomes independent of its mass. This would signal the minimum metallicity possible for a planet and could identify whether such planets form through core-accretion or direct collapse/gravitational instability. 

A different cut-off mass is found for each sample with the transiting spectroscopy, finding the lowest cut-off at about $0.3\,M_{\mathrm{Jup}}$, followed by eclipse spectroscopy at $1.66\,M_{\mathrm{Jup}}$, with direct spectroscopy putting a lower-limit on a mass cut-off at $20\,M_{\mathrm{Jup}}$. As with the slope and intercept of the direct spectroscopy trend, the mass cut-off may be sensitive to a few higher precision, low-mass planet measurements. The Solar System had a poorly constrained mass cut-off with the peak of the posterior distribution at about $0.5\,M_{\mathrm{Jup}}$, but with a long tail towards higher mass, up to the prior edge at $10^5\,M_{\mathrm{Jup}}$. This is due to the fact that the Solar System trend is constrained by only four measurements, maxing out at Jupiter's own mass of $1\,M_{\mathrm{Jup}}$.

With such a wide range of measured mass cut-offs, it is difficult to conclude much about the presence of such a cut-off in the exoplanet population. Future high-precision measurements of planets $>1\,M_{\mathrm{Jup}}$, especially with transit and eclipse spectroscopy, could help detect if such a cut-off exists.

\subsection{C/O Ratio}\label{sec:co}

All C/O ratios measured are consistent with values between 0 and 1, in agreement with the empirical trends from \cite{fu:2025}. The only measurement where the 1$\sigma$ range allows for values of C/O greater than 1 was TOI-5205b, a gas giant around an M-dwarf star with a measured C/O of $1.3\pm0.4$ \citep{canas:2025}. The high C/O ratio is found in two different retrieval analyses, but the spectrum is heavily contaminated by star spots and faculae, perhaps confounding the measurement. On the other hand, the formation of Jovian exoplanets around low-mass stars may proceed with different C/O values than from the vast majority of systems in our sample \citep[e.g.,][]{tabone:2023,sierra:2025}.

With all other planets comfortably between C/O of 0 and 1, it does not appear that carbon-dominated gaseous exoplanets exist, or at least they are exceedingly rare. Some ultra-hot Jupiter measurements have shown relatively high C/O ratios, including WASP-121b from multiple datasets \citep{smith:2024,pelletier:2025,gapp:2025,evans-soma:2025,saha:2025b}, WASP-19b \citep{saha:2025a} and WASP-178b \citep{saha:2025c} (though this is in strong disagreement with a different analysis of the WASP-178b data in \cite{lothringer:2025}). Yet even these relatively high-C/O ratios are constrained to be below 1.0.\footnote{We do not yet include \cite{saha:2025a}, \cite{saha:2025b}, and \cite{saha:2025c} because they are still undergoing peer-review.}

The conclusion that carbon-dominated planets are rare, if they exist at all, was also seen in a more limited sample of \jwst{} observations in \cite{walsh:2025}, and argues strongly against scenarios where gas giants are able to accrete large amounts of carbon-rich gas in the outer protoplanetary disk. Such scenarios are informed by observations of C/O $\gg 1$ gas in protoplanetary disks \citep{bosman:2021b}. The lack of such carbon-rich planets suggests that the reservoir probed by such protoplanetary disk observations is not representative of planet-forming material for the planets in our sample. This is surprising since the mixing should be efficient enough such that no large vertical gradients exist between the disk photosphere and the mid-plane \citep{bosman:2021b}. 

This leads to the conclusion that perhaps the planet-building material is to be found at smaller orbital distances where the gas may be more oxygen-rich, in part perhaps due to the inward drift of icy pebbles, as appears to be the case for several disks \citep[e.g.,][]{bitsch:2022,booth:2023,gasman:2023,schwarz:2024,henning:2024,walsh:2025}. Alternatively, the heavy element composition in the envelope of the planets in our sample could be determined through the accretion of solid, oxygen-rich pebbles or planetesimals, which could be found at larger orbital radii \citep{mordasini:2016,espinoza:2017}. The latter should be the case for planets below about $2-10\,M_{\mathrm{Jup}}$ \citep{mordasini:2016}. It is then somewhat surprisingly to see so many solar-to-sub-solar C/O ratio high-mass exoplanets in the direct spectroscopy sample, which may both be above that mass limit and are presently found at orbital distances consistent with the carbon-rich part of the protoplanetary disk. Possible explanations include that such planets found at large orbital radii may form quickly, before the disk has chemical evolved, or perhaps have formed through gravitational instability \cite{walsh:2025}.

\subsubsection{C/O Ratio and Observing Geometry}\label{sec:co_obs_geo}

In Figure~\ref{fig:met_co_geo}, the transmission spectroscopy sample appears to show systematically lower C/O ratios, with eclipse spectroscopy showing systematically high C/O ratios, relative to the direct spectroscopy C/O ratios, which cluster near the solar value of about 0.55 \citep{asplund:2009}. This qualitative interpretation is supported by basic weighted statistics for these populations, shown in Table~\ref{tab:spec_stats}. The median C/O measured by transit spectroscopy is 0.350, while the median measurement from direct and eclipse spectroscopy is 0.59 and 0.7 respectively, with the standard deviation of the sample of C/O measurements being around 0.2 for each observing geometry (excluding ultra-hot Jupiters). 

We ran a K-S analysis with the C/O ratio measurements, performing both unweighted and weighted K-S tests. In the unweighted K-S tests, transmission spectroscopy was found to be statistically different than both the direct and eclipse spectroscopy samples, with p-values of 0.011 and 0.012, respectively. On the other hand, the direct and eclipse spectroscopy samples were found to be somewhat less statistically different with a more marginal p-value of 0.091. As with metallicity, the entire exoplanet sample was found to be statistically different than the stellar sample from Hypatia with a p-value of $6.57\times10^{-6}$. Again similar to the metallicity analysis, the weighted K-S tests find all samples to be more statistically different, with p-values of $<0.001$.

Only two systems in our sample have measurements in both transit and emission: hot Jupiter HD 189733b and ultra-hot Jupiter WASP-121b. Overall, the measurement at different geometries are consistent with one another. \cite{zhang:2025} observed HD 189733b with \jwst{}/NIRCam in both transit and eclipse, finding C/O ratios of $0.41 \pm ^{+0.13}_{-0.12}$ and $0.43 \pm ^{+0.06}_{-0.05}$, respectively. Meanwhile, \cite{smith:2024,pelletier:2025,evans-soma:2025,pelletier:2025b} each measured WASP-121b in eclipse from the ground and with \jwst{} and \cite{gapp:2025} observed the planet in transit with \jwst{}; in all cases a super-solar C/O of at least 0.7 was measured. The one possible exception is that when \cite{pelletier:2025b} assumed a zero geometric albedo, a more solar C/O ratio of $0.48 \pm ^{+0.14}_{-0.16}$ was obtained.

One other source of bias may be the wavelength range used to measure each planet's spectrum. For example, spectroscopy that only utilized \jwst{}/NIRISS/SOSS (0.6-2.7$\,\mu$m) may not be sensitive to CO, whose fundamental band absorption is $>4\,\mu$m and could underestimate the carbon-content of a planet's atmosphere. However, only two studies in Table~\ref{tab:ExoCompTable} used observations of only \jwst{}/NIRISS/SOSS, WASP-18b \citep{coulombe:2023} and WASP-121b \citep{pelletier:2025b}, and both were in eclipse, not transmission, so this cannot explain the low C/O ratios for planets observed with transit spectroscopy. Considering planets only observed from 3-5$\,\mu$m with either \jwst{}/NIRCam or \jwst{}/NIRSpec/G395H, the median C/O measured is exactly 0.5, consistent with solar (see Table~\ref{tab:spec_stats}). Similarly, studies using ground-based high-resolution spectroscopy show a median C/O of 0.635. We therefore conclude that the instruments or wavelengths used to measure the C/O are not responsible for the low measurements in planets observed with transit spectroscopy.

Below, we show that the measured C/O does not appear to correlate with temperature (Sec.~\ref{sec:co_temp}) or planet mass (Sec.~\ref{sec:co_mass}), suggesting that the low C/O ratio found in the transit spectroscopy sample is not immediately obvious as a sampling bias. 

\subsubsection{C/O-Temperature}\label{sec:co_temp}

Unexpected trends with respect to composition and temperature may be tell-tale signs of retrieval and/or observational biases rather than primordial abundance differences. For example, an ultra-hot Jupiter might exhibit a biased O/H measurement compared to if it were at a lower temperature because thermal dissociation of H$_2$O will make the inference of O/H sensitive to the retrieved T-P profile \citep[e.g.,][]{kitzmann:2018,parmentier:2018,lothringer:2018}. Similarly, at lower temperature, quenching through vertical mixing and/or elevated internal temperatures may bias chemical equilibrium retrievals towards lower C/O ratios and high metallicities to explain the depletion of CH$_4$ that may otherwise be expected to be abundant in equilibrium at such temperatures \citep{visscher:2011}.

\begin{figure}
    \centering
    \includegraphics[width=1.0\linewidth]{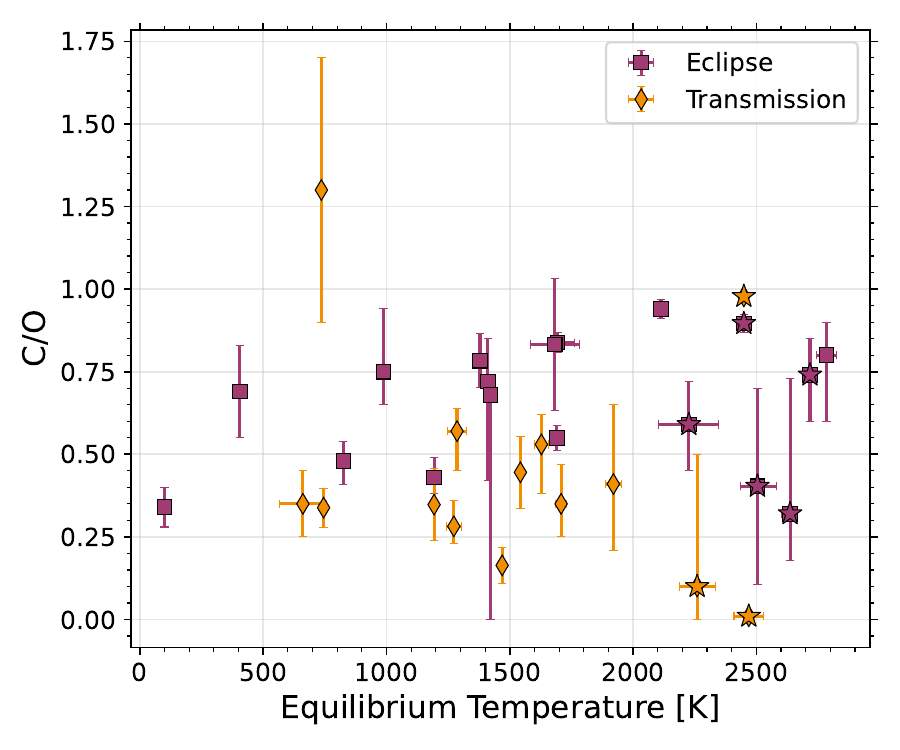}
    \caption{C/O ratio as a function of equilibrium temperature for planets in the eclipse (magenta squares) and transmission (gold diamonds) spectroscopy samples. Ultra-hot Jupiters are labeled as stars.}
    \label{fig:temp_co}
\end{figure}

Figure~\ref{fig:temp_co} shows the correlation between metallicity and C/O ratio for planets measured through eclipse and transmission spectroscopy. We chose to not include planets from direct spectroscopy because their temperatures will be model-dependent and are not currently tracked in the Table~\ref{tab:ExoCompTable}. For both emission and transmission spectroscopy, we find no clear linear trend between C/O ratio and equilibrium temperature with the absolute value of the Pearson correlation coefficients below 0.25 for both. 

We do note that while ultra-hot Jupiters measured in eclipse qualitatively have the same distribution as cooler planets, the ultra-hot Jupiters measured in transmission generally have rather extreme C/O values, either near zero or one. This may be illustrative of the challenges of measuring ultra-hot Jupiter volatile compositions in the presence of thermal dissociation, particularly at the low-pressures probed by transmission spectroscopy.

\subsubsection{Mass-C/O}\label{sec:co_mass}

\begin{figure*}
    \centering
    \includegraphics[width=0.8\linewidth]{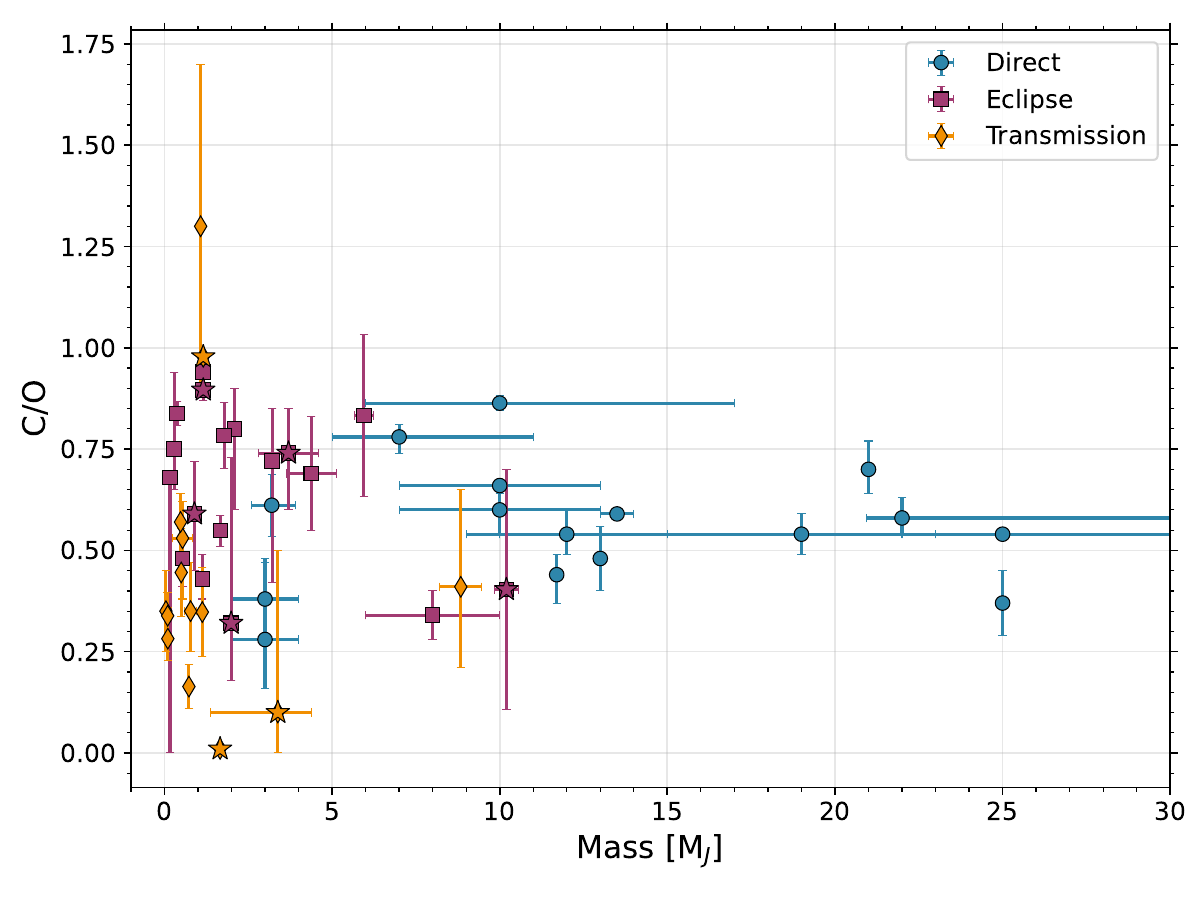}
    \caption{C/O ratio versus planet mass for our sample. Planets observed with direct spectroscopy are shown as blue circles, planets observed with eclipse spectroscopy as magenta squares, and planets observed with transit spectroscopy as gold diamonds. Ultra-hot Jupiters are labeled as stars. No discernible correlation is apparent between the parameters, but the scatter appears to increase at lower mass.}
    \label{fig:co_mass}
\end{figure*}

Since different mass planets are likely to accrete different proportions of gas, rock, and ice, and those reservoirs each are likely to have different proportions of oxygen- and carbon-bearing elements \citep{oberg:2011}, it is possible that the C/O ratio of a planet will change with mass. Figure~\ref{fig:co_mass} shows how C/O varies with mass across our sample.

With a different sample, \cite{hoch:2023} identified a transition between two distinct populations at about $4\,M_{\mathrm{Jup}}$ in the C/O ratio, where lower-mass, mostly transiting planets exhibited a wider range than the higher-mass planets observed with direct spectroscopy, which clustered around solar C/O. As also discussed in Section~\ref{sec:co_met}, we find a similar increase in scatter among the lower-mass transit spectroscopy sample, which show a standard deviation of 0.427 (including ultra-hot Jupiters) or 0.125 (not including ultra-hot Jupiters). This can be compared with the standard deviation in the C/O ratio of the higher-mass direct spectroscopy sample of 0.103.

Weighted correlation coefficients between the C/O ratio and mass do not exceed an absolute value of 0.15 for any sample or subset of the sample, implying no linear correlation between C/O and mass within each sample. As mentioned in Section~\ref{sec:co_obs_geo}, however, K-S tests do indicate a difference between the sample populations. Combining these two tests suggests that the difference in C/O between the populations is not being driven by their difference in mass, but from either different planet properties or biases between the techniques used for the different types of observation.

\subsection{C/O-Metallicity}\label{sec:co_met}

Figure~\ref{fig:met_co} shows the relationship between metallicity and the C/O ratio for the different observing geometries. As for any trend between C/O ratio and metallicity, the maximum Pearson correlation coefficient for the two parameters is only 0.61 for the direct spectroscopy sample. Using a correlation coefficient weighted by the inverse variance of each measurements gives a coefficient of 0.37 for the eclipse spectroscopy followed closely by the direct spectroscopy sample at 0.36. The transmission spectroscopy showed no significant correlation between C/O and metallicity. Only the p-value for the direct spectroscopy sample was found to be statistically significant at $p = 0.0075$.

We note that there is a clear trend in the broader stellar population, with higher metallicity stars showing a higher C/O ratio \citep{hinkel:2014} as a result of galactic chemical evolution. With more measurements from a wider variety of well-characterized host star chemical inventories, one may hope to one day see such an effect of galactic chemical evolution in exoplanet atmosphere abundances.

\begin{figure*}
    \centering
    \includegraphics[width=0.45\linewidth]{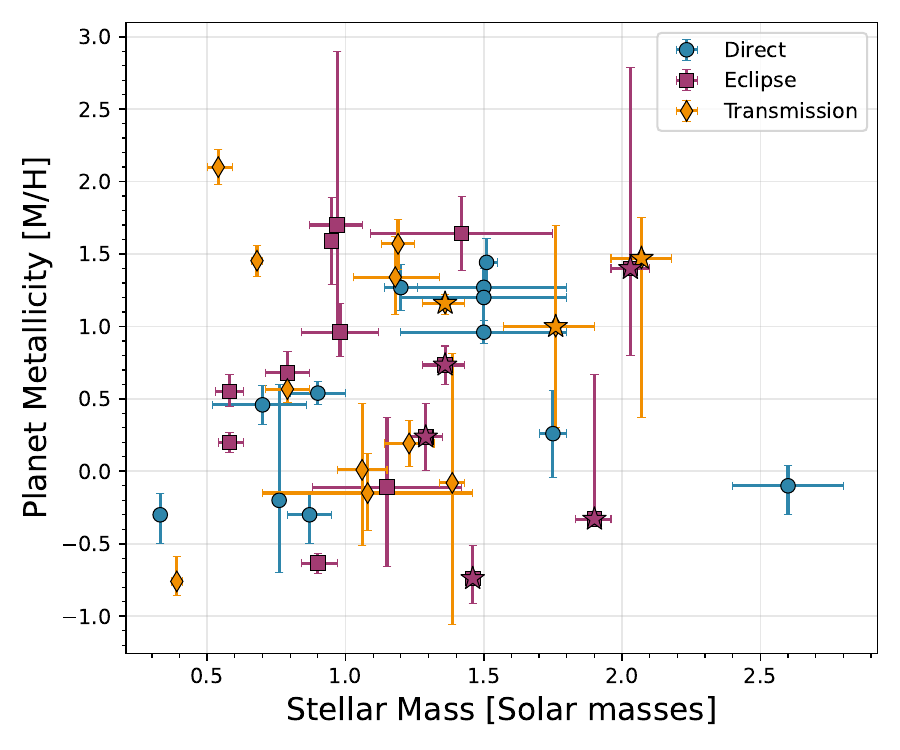}
    \includegraphics[width=0.45\linewidth]{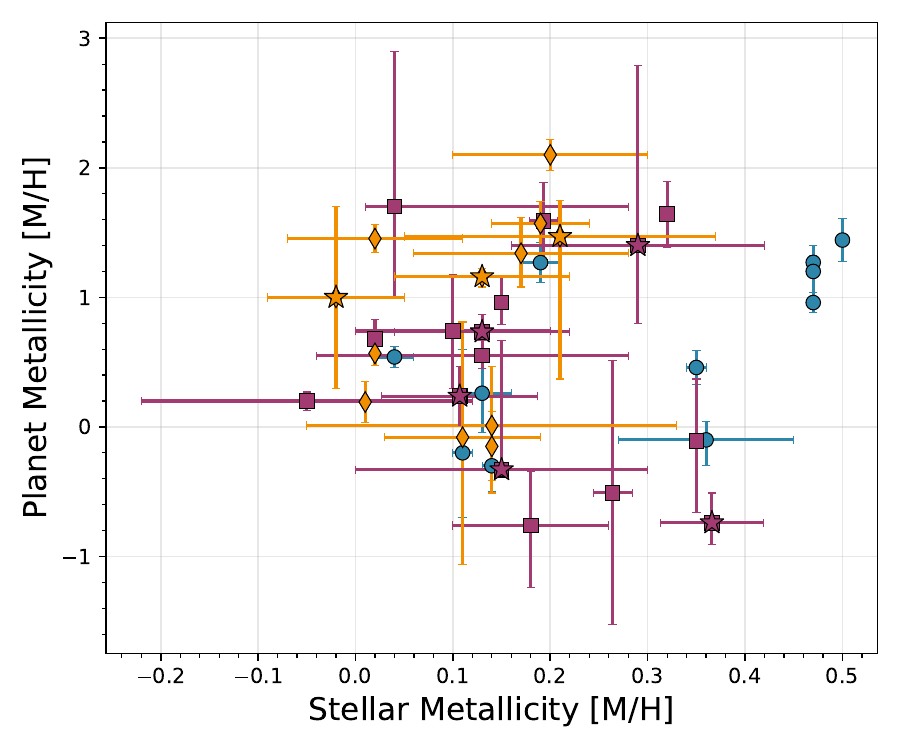}
    \includegraphics[width=0.45\linewidth]{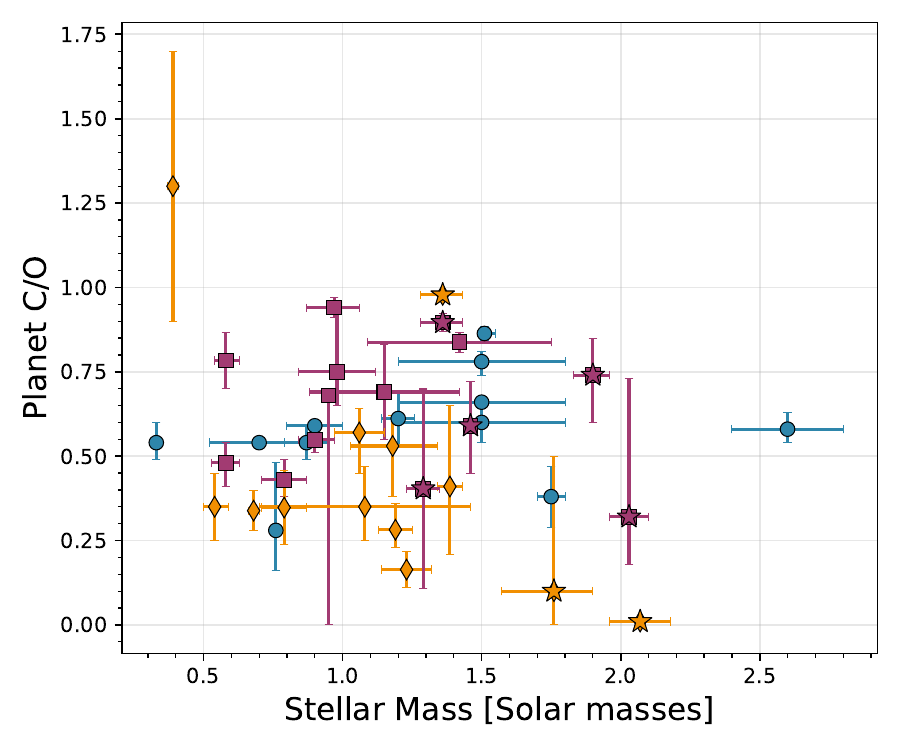}
    \includegraphics[width=0.45\linewidth]{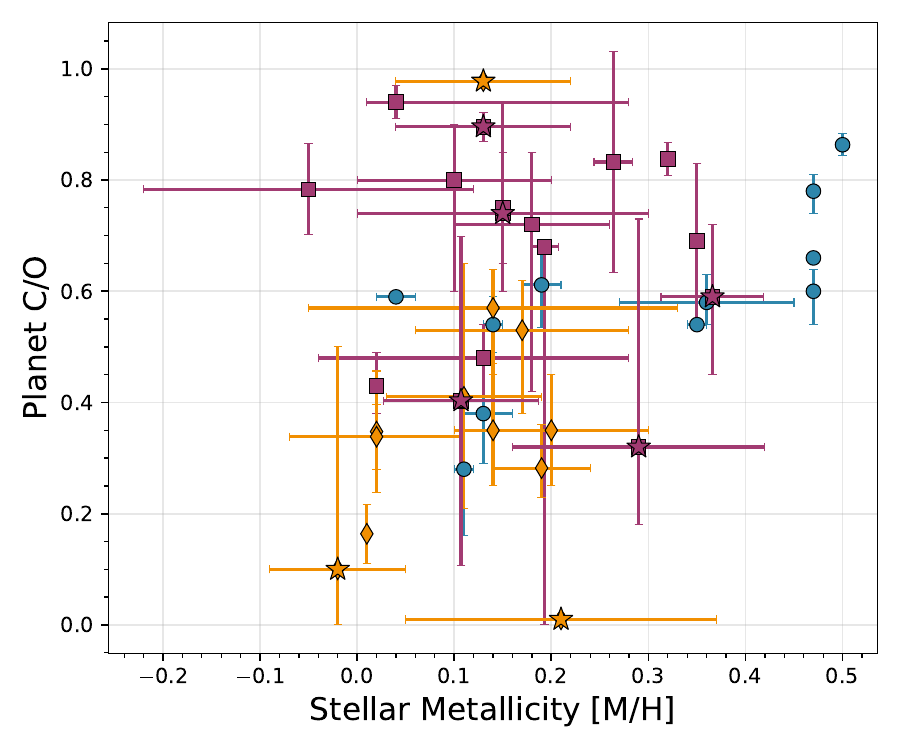}
    \caption{Planet metallicity (top, as given in Table~\ref{tab:ExoCompTable}) and C/O (bottom) versus stellar mass (left) and metallicity (right) as given by the NASA Exoplanet Archive \citep{christiansen:2025}. Planets observed with direct spectroscopy are shown as blue circles, planets observed with eclipse spectroscopy as magenta squares, and planets observed with transit spectroscopy as gold diamonds. Ultra-hot Jupiters are labeled as stars.}
    \label{fig:stellarprop_metallicity}
\end{figure*}

\subsection{Stellar Mass and Metallicity}\label{sec:stellar}

Lastly, we compared our samples of measured planet compositions to host star properties of mass and metallicity. Because the host star's mass and metallicity likely has a fundamental influence on the characteristics of a system's protoplanetary disk, it is hypothesized that the host star properties can influence the resulting exoplanet compositions \citep[e.g.,][]{mordasini:2016}. For example, high mass stars have higher mass protoplanetary disks \citep{andrews:2013}, with correspondingly more material available with which to form planets. Similarly, high-metallicity stars may have high-metallicity disks with more material available to build proto-planet cores or later enrich planetary atmospheres \citep{fischer:2005}, while also affecting the gas-to-solid ratio. Also, as mentioned in Sec.~\ref{sec:co_met}, the C/O of stars positively correlates with their metallicity because of galactic chemical evolution and such a correlation might end up reflected in the exoplanet atmosphere as well. 

Figure~\ref{fig:stellarprop_metallicity} shows the relationship between planet metallicity and the stellar mass (top) and metallicity (bottom), both from the NASA Exoplanet Archive \citep{christiansen:2025}. An unweighted Pearson correlation coefficient shows no significant correlation between these parameters. At $r=0.49$ and a $p$-value of $0.09$, the highest correlation is found between planet and stellar metallicity for the direct spectroscopy sample, though this appears to be driven by the HR 8799 system, for which both the planet and star have high metallicities. Similarly, the only somewhat significant correlation between stellar mass and planet C/O is with the transit spectroscopy sample with $r=-0.48$ and a $p$-value of $0.069$.

\section{Conclusions}
\label{sec:conclude}
In this work, we have introduced \exocomp{}, a Python-based, open-source toolkit to enable the inter-comparison of exoplanet composition retrieval results. The utilities include procedures for converting between different planet abundance (e.g., VMR versus MMR), solar abundance (e.g., \citealt{asplund:2021} versus \citealt{lodders:2025}), metallicity (e.g., O/H versus C/H versus (O+C)/H), and C/O definitions (e.g., modifying C, O, or both), effectively converting planet abundances to absolute abundances. We also provide an alternative method for inferring bulk abundance properties like metallicity, C/O ratio, and Refr./Vol. ratio from free retrieval results using fits to chemical equilibrium solutions.

We then collected all available measurements, selecting those that used observations from \jwst{} and/or 8-meter-class observatories to prioritize data with the currently highest-possible information content. While many studies have pointed out the technical challenges of comparing disparate retrieval results \citep[e.g.,][]{barstow:2020}, being able to more consistently compare the growing library of planet compositions can point the way towards improving and/or verifying exoplanet atmosphere retrievals, while guiding future exoplanet science cases.

We call this collection of measurements the Library of Exoplanet Atmospheric Composition Measurements (LExACoM). We intend to keep the Library up-to-date with new measurements, as our inventory of measurements continues to expand. Authors of new measurements are encouraged to submit new or missing entries by emailing the authors. We archive the table at the time of publication for replication purposes.

To ease future inter-comparison of retrieval results, we provide a set of recommendations for exoplanet abundance reporting is as follows:
\begin{my_itemize}
    \item For the purposes of inferring planet formation information, planetary abundances should be compared to the host star abundances when possible, and not solar abundances.
    \item If quoting abundances relative to solar metallicity, a reference to the solar abundance definition should be indicated.
    \item When quoting a metallicity, C/O ratio, and/or Refr./Vol., the precise definitions of these values and how they are parameterized within the retrieval should be explained.
    \item The preferred or ``take-away" measurements should be identified for a given dataset. This can include weighted sets of constraints when combining different retrieval scenarios \citep[e.g.,][]{lothringer:2025}. If authors do not find it appropriate for a given dataset to (i.e., if complications preclude a consistent interpretation) that should also be stated and respected.
\end{my_itemize}

Using the Library, we search for trends with respect to metallicity, C/O ratio, observing geometry, and external properties like the planet mass and stellar metallicity. All calculations in this paper are replicated in Jupyter Notebooks as part of the documentation for \exocomp{} (see Footnotes). We find:
\begin{my_itemize}
    \item Measured metallicities vary between 0.1 and $100\times$ solar, with an more scatter than expected in metallicity at and above $1\,M_{\mathrm{Jup}}$.
    \item Elevated metallicity across the exoplanet population with respect to T-dwarf \citep{zalesky:2022} and stellar populations \citep{hinkel:2014}.
    \item All C/O ratios are consistent within 1$\sigma$ with a range between 0 and 1.
    \item A systematically low average C/O ratio for planets measured by transmission spectroscopy compared to those measured with eclipse and direct spectroscopy.
    \item No discernible trend between metallicity and C/O ratio.
    \item C/O ratio and metallicity do not clearly correlate with temperature, indicating that there is not a clear bias in retrievals due to temperature-dependent processes like CH$_4$-depletion and H$_2$O thermal dissociation.
    \item The canonical mass-metallicity relationship is seen in the transit and eclipse spectroscopy sample.
    \begin{my_itemize}
        \item While the slope of the mass-metallicity trend agrees well between the transit and eclipse sample, they are offset in the intercept and mass cut-off.
        \item The Solar System's mass-metallicity trend appears somewhat shallower than the transit and eclipse sample. 
        \item The mass-metallicity relationship fit to the transit sample can replicate the metallicity of Jupiter, Uranus, and Neptune, and only slightly over-predicts the metallicity of Saturn.
        \item The slope of the \emph{atmospheric} mass-metallicity trend found here is steeper and at lower metallicities than the corresponding slope for the \emph{bulk} metallicity found from interior modeling \citep{thorngren:2016}, perhaps implying a decreasing core mass fraction and/or increasing mixing as a function of decreasing mass.
        \end{my_itemize}
    \item No correlation is found between C/O ratio and mass.
    \item No correlation is found between planet metallicity or C/O and the stellar mass or metallicity.
\end{my_itemize}

We hope that the tools available in \exocomp{} and the data in the Library of Exoplanet Atmospheric Composition Measurements can enhance the science return of the wide-array of space- and ground-based exoplanet science being undertaken by the community. Future enhancements would include expanding the Library to include free retrieval results, enabling an inventory of individual atomic and molecular abundances across the exoplanet population. Similarly, an expansion of stellar abundance measurements in the Library would enhance the power of planet formation inferences. With consistent, population-level data, we can begin to tackle some of the most complicated questions in exoplanet science, including the connection between exoplanet composition and planet formation.

\begin{acknowledgments}

We thank the scientific reviewer and data editor for their helpful reviews, which improved the paper. We acknowledge the use of the ExoAtmospheres database during the preparation of this work. The research shown here acknowledges use of the Hypatia Catalog Database, an online compilation of stellar abundance data as described in \cite{hinkel:2014}, which was supported by NASA's Nexus for Exoplanet System Science (NExSS) research coordination network and the Vanderbilt Initiative in Data-Intensive Astrophysics (VIDA). This research has made use of NASA's Astrophysics Data System Bibliographic Services. This research has made use of the NASA Exoplanet Archive, which is operated by the California Institute of Technology, under contract with the National Aeronautics and Space Administration under the Exoplanet Exploration Program. NL acknowledges the support of the UD Annie Jump Cannon Fund PHYS462112. The version of \exocomp{} used in this paper is archived in Zenodo under \cite{exocomp_zenodo}.
\end{acknowledgments}

\facilities{JWST, VLT:Melipal, VLTI, Gemini:South, Keck I}

\software{\texttt{astropy} \citep{astropy:2018},  
          \texttt{Cloudy} \citep{ferland:1998,gunasekera:2023},
          \texttt{EasyCHEM} \citep{lei:2024},
          \texttt{Dynesty} \citep{fer08,fer09,skilling:2004,koposov:2024}
          }

\bibliography{ms_new}{}
\bibliographystyle{aasjournalv7}

\end{document}